\documentclass[aps,pra,twocolumn,showpacs,notitlepage,superscriptaddress,letterpaper,amsmath,amssymb]{revtex4-2}

\usepackage{graphicx}
\usepackage{dcolumn}
\usepackage{bm}
\usepackage{braket}
\usepackage{lipsum}
\usepackage{mathtools}
\usepackage{hyperref}

\begin{document}


\title{Non-classical microwave-optical photon pair generation with a chip-scale transducer}


\author{Srujan Meesala}
\thanks{These authors contributed equally}
\affiliation{Kavli Nanoscience Institute and Thomas J. Watson, Sr., Laboratory of Applied Physics, California Institute of Technology, Pasadena, California 91125, USA}
\affiliation{Institute for Quantum Information and Matter, California Institute of Technology, Pasadena, California 91125, USA}
\author{Steven Wood}
\thanks{These authors contributed equally}
\affiliation{Kavli Nanoscience Institute and Thomas J. Watson, Sr., Laboratory of Applied Physics, California Institute of Technology, Pasadena, California 91125, USA}
\affiliation{Institute for Quantum Information and Matter, California Institute of Technology, Pasadena, California 91125, USA}
\author{David Lake}
\thanks{These authors contributed equally}
\affiliation{Kavli Nanoscience Institute and Thomas J. Watson, Sr., Laboratory of Applied Physics, California Institute of Technology, Pasadena, California 91125, USA}
\affiliation{Institute for Quantum Information and Matter, California Institute of Technology, Pasadena, California 91125, USA}
\author{Piero Chiappina}
\affiliation{Kavli Nanoscience Institute and Thomas J. Watson, Sr., Laboratory of Applied Physics, California Institute of Technology, Pasadena, California 91125, USA}
\affiliation{Institute for Quantum Information and Matter, California Institute of Technology, Pasadena, California 91125, USA}

\author{Changchun Zhong}
\affiliation{Pritzker School of Molecular Engineering, The University of Chicago, Chicago, IL 60637, USA}

\author{Andrew D. Beyer}
\affiliation{Jet Propulsion Laboratory, California Institute of Technology, 4800 Oak Grove Dr, Pasadena, California 91109, USA}

\author{Matthew D. Shaw}
\affiliation{Jet Propulsion Laboratory, California Institute of Technology, 4800 Oak Grove Dr, Pasadena, California 91109, USA}

\author{Liang Jiang}
\affiliation{Pritzker School of Molecular Engineering, The University of Chicago, Chicago, IL 60637, USA}

\author{Oskar~Painter}
\email{opainter@caltech.edu}
\homepage{http://copilot.caltech.edu}
\affiliation{Kavli Nanoscience Institute and Thomas J. Watson, Sr., Laboratory of Applied Physics, California Institute of Technology, Pasadena, California 91125, USA}
\affiliation{Institute for Quantum Information and Matter, California Institute of Technology, Pasadena, California 91125, USA}
\affiliation{Center for Quantum Computing, Amazon Web Services, Pasadena, California 91125, USA}

\date{\today}

\maketitle

\textbf{
Modern computing and communication technologies such as supercomputers and the internet are based on optically connected networks of microwave frequency information processors \cite{ChengDataCenters18, Thraskias2018}. In recent years, an analogous architecture has emerged for quantum networks with optically distributed entanglement \cite{Cirac1997,Kimble2008} between remote superconducting quantum processors, a leading platform for quantum computing \cite{Kjaergaard2020, arute2019quantum, Wu2021}. Here we report an important milestone towards such networks by observing non-classical correlations between photons in an optical link and a superconducting electrical circuit. We generate such states of light through a spontaneous parametric down-conversion (SPDC) process in a chip-scale piezo-optomechanical transducer \cite{Zhong2020FreqBin}. The non-classical nature of the emitted light is verified by observing anti-bunching in the microwave state conditioned on detection of an optical photon. Such a transducer can be readily connected to a superconducting quantum processor, and serve as a key building block for optical quantum networks of microwave frequency qubits.
}

Networks of remotely situated qubits \cite{Cirac1997, Kimble2008} are essential to harness quantum correlations for long-distance secure communication \cite{Briegel1998, Duan2001}, distributed quantum computation \cite{Cirac1999, monroe2014large} and precision measurements \cite{Komar2014, Khabiboulline2019}. Optical photons are naturally suited to act as flying qubits over room temperature links and distribute entanglement in such networks \cite{Chen2021}. Quantum optical networks with few nodes have been realized with systems such as atoms \cite{Welte2021, Yu2020}, quantum dots \cite{Stockill2017, Delteil2016}, trapped ions \cite{Hucul2015}, and color centers \cite{Bernien2013}, which naturally possess optical frequency transitions between their internal energy levels. In parallel developments, superconducting circuits based on Josephson junctions have emerged as a leading platform for quantum information processing with the ability to realize entangled states of many qubits in microwave frequency circuits \cite{Kjaergaard2020, arute2019quantum, Wu2021}. However, superconducting qubits do not possess a natural, coherent interface with optical photons. This limitation has motivated recent efforts to develop transducers capable of generating quantum correlations between optical photons and microwave frequency qubits. While schemes to produce such states are fundamentally well-understood, preserving fragile quantum correlations during the transduction process has not been possible so far in a wide variety of physical platforms \cite{HongTangTransductionReview, JILA_Nondestructive, HTang_Transduction_EO_AlN, HTang_Transduction_EO_LN, Magnon_Transduction, Faraon_Transduction}. This roadblock is primarily due to technical challenges posed by the vast difference in energy scales between microwave and optical photons.

Here we demonstrate non-classical microwave-optical photon pairs from such a transducer. We use a piezo-optomechanical device in which an acoustic mode acts as an intermediary between microwave and optical fields. First, a pump laser pulse generates photon-phonon pairs in an optomechanical cavity via an SPDC process. Subsequently, a strong piezoelectric interaction converts the phonon into a microwave photon. To characterize the photon pairs emitted by the transducer, we simultaneously perform single photon detection of the optical component and heterodyne detection of the microwave component. For experimental trials in which single optical photons are detected, we compute the normalized second order intensity correlation function of the conditional microwave state. Observation of a value below unity for this function constitutes direct verification of non-classical statistics of the photon pairs. Compared with a recent demonstration of microwave-optical two-mode squeezing \cite{sahu2023}, our experiment demonstrates the requisite techniques to implement the well-known DLCZ protocol \cite{Duan2001}, which uses detection of optical photons to herald entanglement between distant nodes in quantum networks.

Our result is primarily enabled by fabricating nearly all circuit components of a chip-scale piezo-optomechanical transducer from niobium nitride (NbN), a superconductor in which quasiparticles (QPs) generated by optical absorption relax on the timescale of a few nanoseconds \cite{NbNQP}. Compared with our previous work in which we directly coupled such a transducer to a transmon qubit on the same chip \cite{Mirhosseini2020}, using a circuit with fast QP relaxation allows us to mitigate the effects of parasitic absorption of stray optical pump light by superconducting components of the transducer. In particular, we add well below one quantum of noise to the circuit during transduction while increasing the rate of probabilistic optical detection events by two orders of magnitude. While superconductors with fast QP relaxation have been employed in piezo-optomechanical transducers recently \cite{Jiang2022opticallyheralded, Weaver2022}, these demonstrations were limited to classical states of light. Our device design and optical detection setup provide the efficiency and noise levels required to observe non-classical correlations.

\begin{figure*}
\includegraphics[width=\textwidth,scale=1]{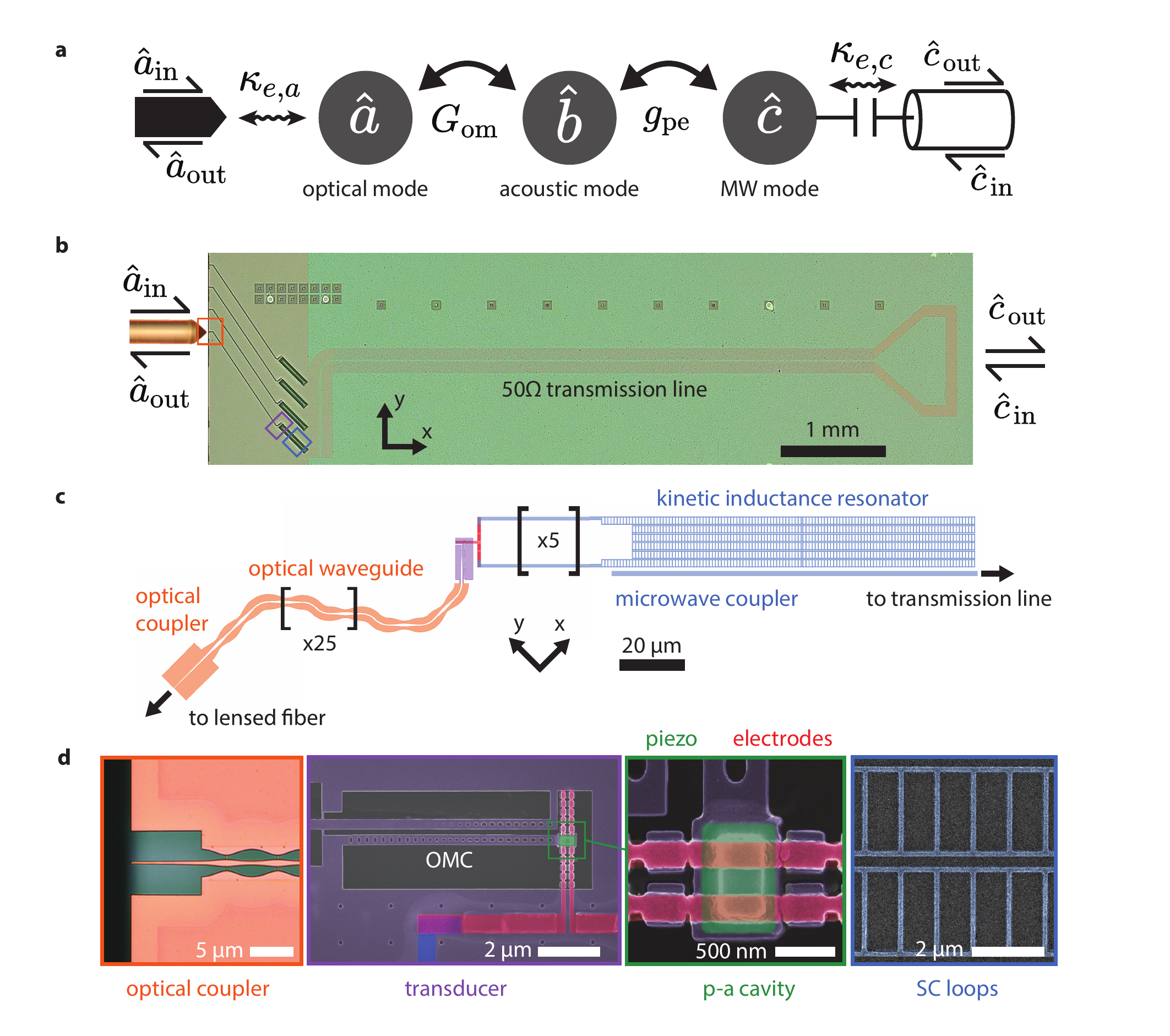}
\caption{\textbf{Quantum transducer}. \textbf{a.} Transducer mode schematic indicating optical ($\hat{a}$), acoustic ($\hat{b}$) and microwave ($\hat{c}$) modes along with interaction rates for optomechanical coupling ($G_{\mathrm{om}}$) and piezoelectric coupling ($g_{\mathrm{pe}}$), respectively. The optical and microwave modes are coupled to waveguides with external coupling rates, $\kappa_{e,a}$ and $\kappa_{e,c}$, respectively. The input and output modes in the optical and microwave waveguides are denoted by $\hat{a}_{\mathrm{in}}, \hat{a}_{\mathrm{out}}$, $\hat{c}_{\mathrm{in}}, \hat{c}_{\mathrm{out}}$, respectively. \textbf{b.} Micrograph of transducer device showing optical access via lensed fiber on the left and microwave access via 50$\Omega$ transmission line on the right. \textbf{c.} Device schematic. From left to right, suspended optical waveguide (orange) leading to the piezo-optomechanical transducer (purple) whose electrical terminals (red) are connected to a microwave kinetic inductance resonator (blue). \textbf{d.} Micrographs of various components of the transducer. From left to right: optical micrograph of the coupler section at the end of the optical waveguide; scanning electron micrograph of the transducer indicating the silicon optomechanical crystal (OMC) cavity in purple; close-up of the piezo-acoustic (p-a) cavity highlighting the piezoelectric material in green and electrodes in red; scanning electron micrograph of the superconducting (SC) loops in the meandering ladder trace of the kinetic inductance resonator.
}
\label{fig1:wide}
\end{figure*}

Figure~\ref{fig1:wide}a shows a conceptual schematic highlighting the resonator modes involved in our transduction experiment. Interaction between an optical mode, $\hat{a}$ and a microwave-frequency acoustic mode, $\hat{b}$ is mediated by a pump laser driving an optomechanical cavity in the resolved sideband regime. Simultaneously, the acoustic mode is resonantly coupled to a microwave-frequency electrical mode, $\hat{c}$ via the piezoelectric effect. We can write the Hamiltonian for this system as
\begin{equation}
    \hat{H}/\hbar = -\Delta_a\hat{a}^\dag\hat{a} + \omega_b\hat{b}^\dag\hat{b} + \omega_c\hat{c}^\dag\hat{c} + \hat{H}_{\mathrm{om}}/\hbar + \hat{H}_{\mathrm{pe}}/\hbar
    \label{eq:ham}
\end{equation}
Here $\omega_{b}, \omega_{c}$ are the frequencies of the modes $\hat{b}, \hat{c}$ respectively and $\hat{H}_{\mathrm{om}}, \hat{H}_{\mathrm{pe}}$ are the optomechanical and piezoelectric interaction Hamiltonians described in more detail below. $\Delta_a = \omega_{\mathrm{p}} - \omega_a$ is the difference between the frequency of the optical pump, $\omega_{\mathrm{p}}$ and that of the optical mode, $\omega_a$. Setting the frequency of the pump laser to be red ($\Delta_a<0$) or blue detuned  ($\Delta_a>0$) with respect to the optical cavity resonance allows us to select either beam-splitter or two-mode squeezing interactions, respectively \cite{aspelmeyer2014cavity}. The first setting can be used to transfer states between the acoustic mode and the optical mode when the transducer is operated as a frequency converter \cite{Mirhosseini2020, Weaver2022, ASN_Transduction}. In this work, we use the latter setting to generate non-classical pairs of optical photons and acoustic phonons in an SPDC process. This choice is motivated by recent proposals for heralded remote entanglement generation which indicate that operation in SPDC mode relaxes the efficiency requirements for piezo-optomechanical transducers \cite{Zhong2020TimeBin, Zhong2020FreqBin, Zhong2022Capacity}. In this setting, we have $\hat{H}_{\mathrm{om}}/\hbar = -G_{\mathrm{om}}(t)(\hat{a}^\dag\hat{b}^\dag + \hat{a}\hat{b})$. The time-dependent optomechanical coupling rate $G_{\mathrm{om}}(t) = \sqrt{n_a(t)} g_{\mathrm{om}}$ is controlled parametrically via the intra-cavity photon population $n_a(t)$ due to the detuned pump laser. Here $g_{\mathrm{om}}$ denotes the optomechanical coupling rate at the single optical photon and acoustic phonon level. The piezoelectric interaction is described by the beam-splitter Hamiltonian $\hat{H}_{\mathrm{pe}}/\hbar = -g_{\mathrm{pe}}(\hat{b}^\dag\hat{c} + \hat{b}\hat{c}^\dag)$. Here $g_{\mathrm{pe}}$ denotes the piezoelectric coupling rate at the single microwave photon and acoustic phonon level. This interaction can be used to map the acoustic component of the optomechanical two-mode squeezed state onto the microwave electrical mode. In the absence of any added noise, the joint state of the modes, $\hat{a}, \hat{c}$ can be described in the photon number basis by the wavefunction $\ket{\psi} = \ket{00} + \sqrt{p}\ket{11} + p\ket{22} + O(p^{3/2})$. For a weak pump field, the higher order terms may be neglected, and the transducer emits single optical and microwave photons in pairs with probability, $p \ll 1$. Detection of a single optical photon can then be used to conditionally prepare or herald a single microwave photon.

Figure~\ref{fig1:wide}b shows the physical schematic of our device, which consists of a half-wavelength superconducting kinetic inductance resonator \cite{Zmuidzinas2012} coupled to a piezo-optomechanical transducer. The transducer itself comprises a half-wavelength aluminum nitride (AlN) piezo-acoustic cavity attached to a silicon optomechanical crystal (OMC) resonator via an acoustic waveguide \cite{Mirhosseini2020}. We achieve piezoelectric coupling in our system by using the two end terminals of the microwave resonator as electrical leads over the AlN section of the transducer. The microwave resonator is patterned in a disordered, thin film of NbN in a meandering ladder geometry. The inclusion of closed loops in the thin superconducting film, as shown in Fig.~\ref{fig1:wide}c, allows for tuning of kinetic inductance, and hence tuning of the frequency of the microwave resonator via an external magnetic field \cite{Xu2019}. Using narrow superconducting wires in a meandering geometry results in high impedance, which helps us achieve strong piezoelectric coupling, $g_{\mathrm{pe}}$ with a small acoustic mode volume \cite{SI}. This design strategy is followed to maximize the fraction of acoustic energy in silicon, which has the lowest acoustic loss in our material stack. The NbN resonator is capacitively coupled to a 50$\Omega$ transmission line (not shown in schematic) to facilitate microwave spectroscopy of the transducer. Likewise, the optical cavity is coupled to a waveguide, which terminates in a tapered coupler at the edge of the chip allowing for efficient coupling to a lensed optical fiber. The layout of our device is chosen to reduce optical flux from stray pump light at the microwave resonator. We use an $\sim$1~mm long optical waveguide to physically separate the circuit section of the transducer from the optical coupler, where there is significant scattering of pump light. Additionally, we use extended electrical terminals to physically separate the optically sensitive current anti-node of the kinetic inductance resonator from the OMC, and reduce the impact of local pump scattering.

\begin{figure}
\includegraphics[width=\columnwidth,scale=1]{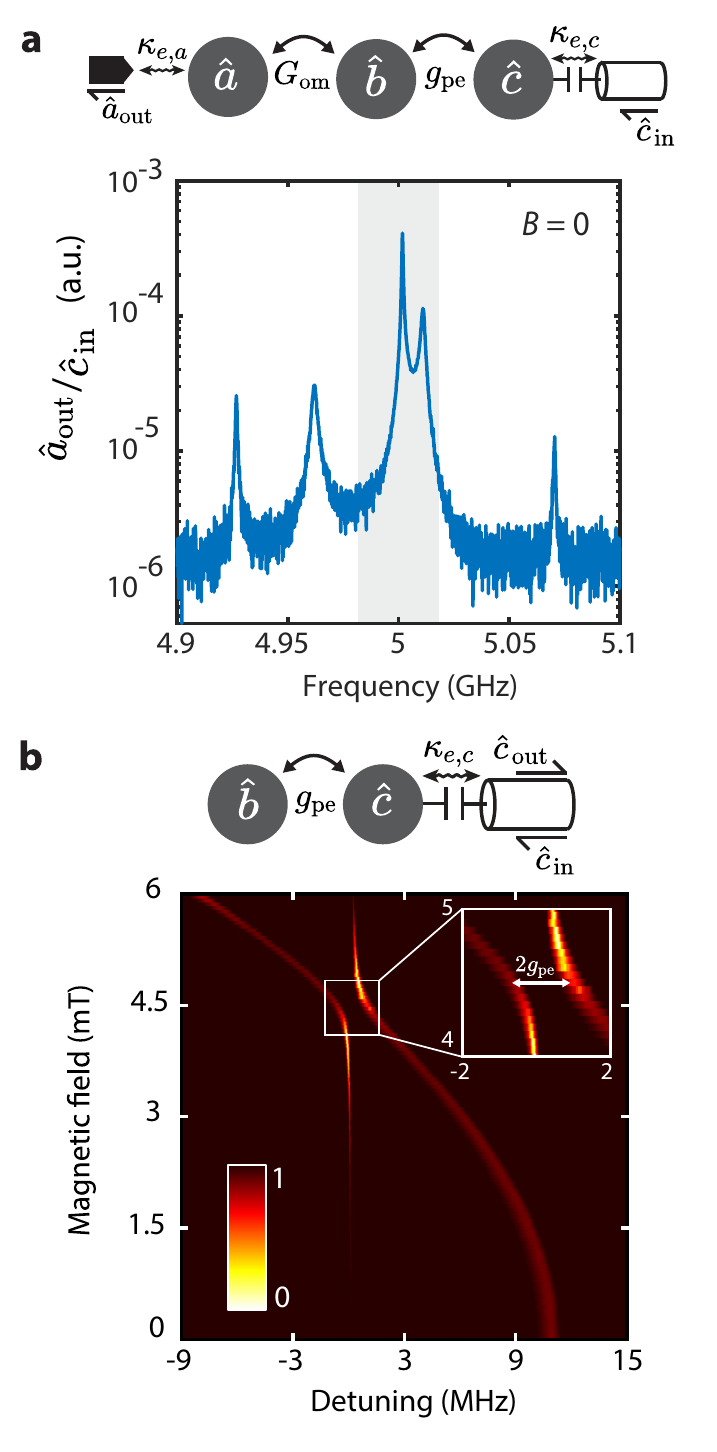}
\caption{\textbf{Optical and microwave spectroscopy}. \textbf{a.} Continuous wave transduction spectrum measured using a vector network analyzer (VNA) at zero magnetic field. The microwave resonator is excited via the input mode, $\hat{c}_{\mathrm{in}}$, and the optical output mode, $\hat{a}_{\mathrm{out}}$ is electrically detected via its microwave frequency beat note with the optical pump on a high speed photodetector. \textbf{b.} Microwave reflection spectrum of the transducer in the frequency range shown by the gray shaded window in panel \textbf{a} probed as a function of external magnetic field. The horizontal axis refers to detuning from the acoustic mode at zero magnetic field. Inset shows a close-up of the anti-crossing between the microwave-frequency electrical and acoustic modes revealing a minimum mode splitting, $2g_{\mathrm{pe}}/2\pi$ = 1.6MHz. 
\label{fig2:wide}}
\end{figure}

Our experiments are carried out by mounting the transducer chip on the mixing plate of a dilution refrigerator. We initially perform optical and microwave spectroscopy to identify the frequencies of the internal transducer modes as well as the optomechanical and piezoelectric coupling rates. For the device used in the experiments that follow, we found an optical resonance at a wavelength $\lambda=1561.3$ nm with critical coupling to the external waveguide, $\kappa_{e,a}/2\pi=\kappa_{i,a}/2\pi=650$ MHz. The subscripts $i,e$ refer to linewidths due to coupling to internal and external baths, respectively. We identify the hybridized microwave-frequency electrical and acoustic modes supported by the integrated transducer-resonator system in Fig.~\ref{fig2:wide}a. In this measurement, we use a vector network analyzer (VNA) to electrically excite the microwave resonator via the microwave input port while optically pumping the optomechanical cavity of the transducer with a laser tuned to the blue side of the optical resonance. The optical pump reflected from the OMC and an optical sideband generated due to transduction of the input microwave signal are together sent to a high-speed photodetector whose output is connected to the detection port of the VNA. The magnitude of signal generated in this VNA spectrum by each of the hybridized microwave-frequency modes of the transducer-resonator system is proportional to the transduction efficiency of the corresponding mode. For the mode at a frequency of 5.001GHz with the highest transduction efficiency and narrowest linewidth, we perform pump power dependent optomechanical spectroscopy and measure an optomechanical coupling rate, $g_{\mathrm{om}}/2\pi$ = 270 kHz. In Fig.~\ref{fig2:wide}b, we show the corresponding microwave reflection spectrum of this mode as we sweep a magnetic field applied perpendicular to the sample. We observe an anti-crossing between the microwave electrical resonator mode, which is identified through its characteristic quadratic tuning response in an external magnetic field \cite{SI}, and the acoustic mode of the transducer. The minimum frequency splitting between these two modes allows us to estimate the piezoelectric coupling rate, $g_{\mathrm{pe}}/2\pi$ = 800 kHz. The independent linewidths of the modes are measured far from the anti-crossing and are found to be $\kappa_{i,b}/2\pi$ = 150 kHz for the microwave acoustic mode and $\kappa_{e,c}/2\pi$ = 1.2 MHz, $\kappa_{i,c}/2\pi$ = 550 kHz for the microwave electrical resonator mode. For the experiments that follow, we set the external magnetic field at the value corresponding to the minimum mode splitting of $2g_{\mathrm{pe}}/2\pi$ = 1.6 MHz where both modes are maximally hybridized. This corresponds to the condition, $\omega_b=\omega_c$ in Eq.~(\ref{eq:ham}). In this setting, we define the hybridized electromechanical modes, $\hat{c}_\pm = (\hat{b}\pm\hat{c})/\sqrt{2}$ with frequencies, $\omega_\pm = \omega_c \pm g_{\mathrm{pe}}$ respectively. Even though the transducer supports other microwave-frequency acoustic modes as shown in Fig.~\ref{fig2:wide}a, these are far detuned from the modes of interest, $\hat{c}_\pm$, relative to the coupling rates, $g_{\mathrm{pe}}$ and $G_{\mathrm{om}}$. As a result, we expect the Hamiltonian in Eq.~(\ref{eq:ham}) to provide a sufficiently accurate description of our system.

\begin{figure}[!h]
\includegraphics[width=\columnwidth,scale=1]{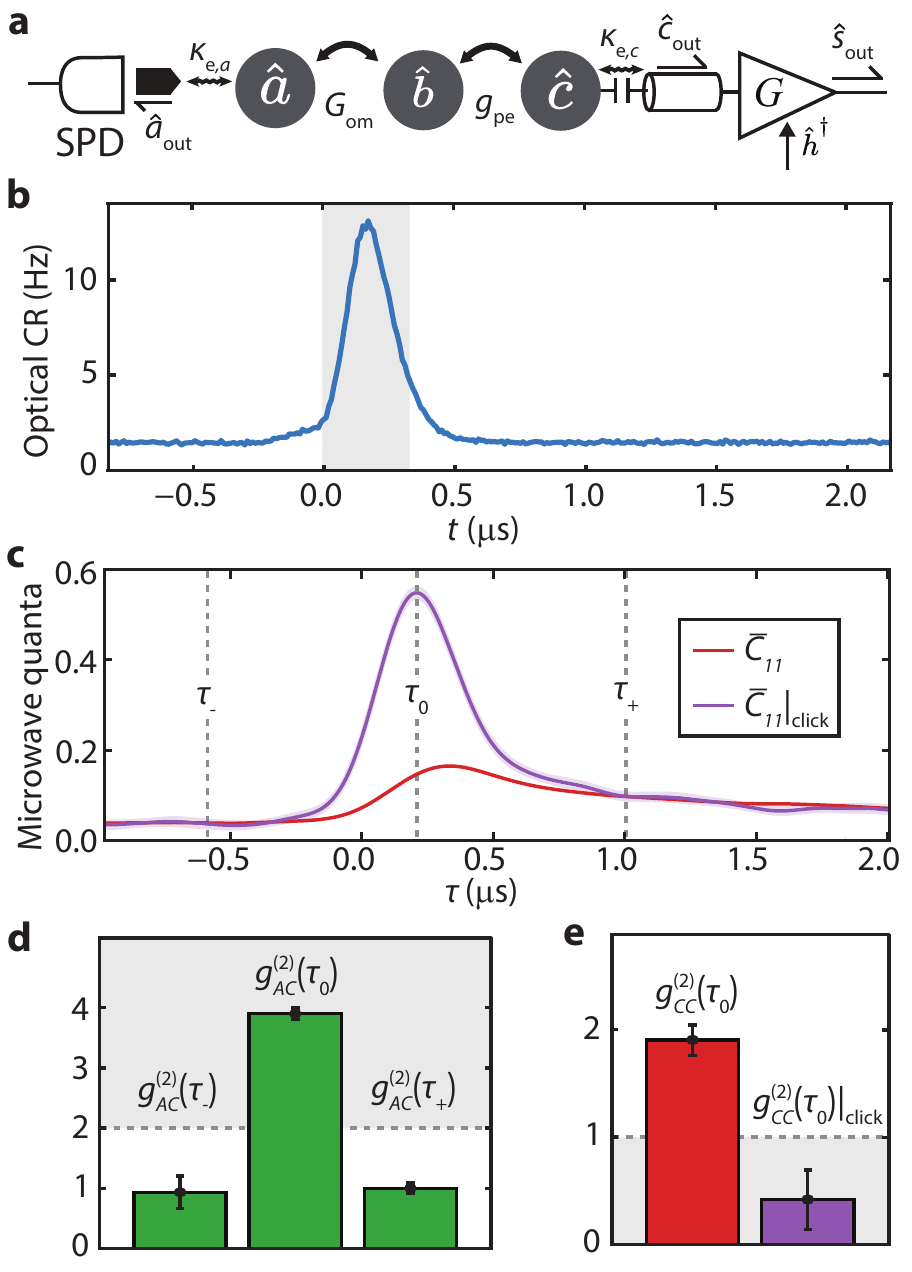}
\caption{\textbf{Microwave-optical cross-correlations.} \textbf{a.} Schematic of the microwave-optical cross-correlation measurement. The optical output is directed to a single photon detector (SPD) and the microwave output is directed to a heterodyne detection setup with gain, $G$ and added noise, $\hat{h}^\dag$. \textbf{b.} Time trace of average optical photon count rate (CR) registered on the SPD. Shaded vertical window indicates the gate duration defining the temporal mode, $\hat{A}$ used to perform conditional microwave readout. \textbf{c.} Microwave quanta in the temporal mode, $\hat{C}$ at the transducer output port as a function of delay, $\tau$ the from the center of the optical gating window. Red trace corresponds to unconditional microwave readout and purple trace corresponds to microwave readout conditioned on  an optical click. Shaded region about the trace indicates a confidence interval spanning two standard deviations about the mean. \textbf{d.} Normalized microwave-optical intensity cross-correlation function at delays $\tau_0$, $\tau_\pm$ indicated in panel \textbf{b}. Dashed line indicates expected classical upper bound for thermal states. \textbf{e.} Normalized second order intensity correlation function of the unconditional microwave state (red) and the microwave state conditioned an optical click (purple). Dashed line indicates classical lower bound.}
\label{fig3:wide}
\end{figure}

We operate the transducer in SPDC mode by exciting it with optical pump pulses at the blue optomechanical sideband of the optical cavity ($\Delta_{\mathrm{a}} = (\omega_+ + \omega_-)/2$). In the experiments described below, Gaussian pump pulses of full-width at half-maximum (FWHM) duration, $T_{\mathrm{p}}$ = 160 ns with a peak power corresponding to intra-cavity optical photon occupation, $n_a$ = 0.8 were sent to the device at a repetition rate of 50 kHz. Each pulse represents an experimental trial with a finite probability of generating a microwave-optical photon pair. Figure~\ref{fig3:wide}a shows a schematic of the setup used to characterize the optical and microwave emission from the transducer under these conditions. The optical emission in the mode, $\hat{a}_{\mathrm{out}}$, is sent to a superconducting nanowire single photon detector (SNSPD) after passing through a Fabry-Perot filter setup to suppress the pump pulses reflected by the transducer. In Fig.~\ref{fig3:wide}b, we show the optical photon flux at the SNSPD obtained by averaging single photon `clicks' over multiple trials of the experiment. Gating the detection events in a time window of duration, 2$T_{\mathrm{p}}=320$ ns shown by the gray shaded region, we obtain a click probability, $p_{\mathrm{click}}=2.7\times10^{-6}$, which leads to an optical heralding rate, $R_{\mathrm{click}}=0.14$ Hz. A detailed account of various contributions to this heralding rate is provided in Ref.~\cite{SI}.

In parallel with optical photon detection, the microwave emission from the transducer in the output mode, $\hat{c}_{\mathrm{out}}$ is sent to a heterodyne detection setup as shown in Fig.~\ref{fig3:wide}a. Using the input-output formalism for a phase-insensitive amplifier \cite{Clerk2010} in the limit $G \gg 1$, we write the output of this setup as $\hat{s}_{\mathrm{out}} \approx \sqrt{G}(\hat{c}_{\mathrm{out}} + \hat{h}^{\dag})$, where $\hat{h}^{\dag}$ is the noise mode added by the amplifier. Emission from the transducer modes $\hat{c}_\pm$ is concentrated in a small bandwidth about the frequencies $\omega_\pm$ within $\hat{s}_\text{out}$. We isolate  this signal by integrating the recorded heterodyne voltage signal with a matched emission envelope function $f(t)$ as detailed in the supplementary information \cite{SI}. This corresponds to a measurement of the quadratures of the temporal mode $\hat{S}(t) := \int \hat{s}_\text{out}(t+t')f^{*}(t') dt'$ at the output of the heterodyne detection setup. Similarly, we define $\hat{C}(t):= \int \hat{c}_\text{out}(t+t')f^{*}(t') dt'$ and $\hat{H}(t):= \int \hat{h}_\text{out}(t+t')f^{*}(t') dt'$ to be temporal modes corresponding to emission referred to the output port of the transducer device and amplifier noise. As a consequence of the large bandwidth of the pump pulse compared to the mode splitting ($\eta_\text{TB}/T_{\mathrm{p}}$ = 2.75 MHz $> 2g_\text{pe}/2\pi = 1.6$ MHz, where $\eta_\text{TB}=0.44$ is the time-bandwidth product of a Gaussian pulse) we cannot individually resolve emission coming from $\hat{c}_\pm$. With this in mind, $f(t)$ is constructed to capture emission from both $\hat{c}_\pm$, and consequently the temporal modes have large spectral overlap with both hybridized modes $\hat{c}_\pm$. By measuring the heterodyne voltage signal in experimental trials carried in the presence (absence) of optical pump pulses, we collect complex-valued voltage samples of $\hat{S}$ ($\hat{H}^\dag$). These voltage samples are then used to calculate the moments of the microwave field, $\bar{C}_{mn} = \langle \hat{C}^{\dag m} \hat{C}^n \rangle$ by taking an ensemble average over the experimental trials and inverting the amplifier input-output relations \cite{Eichler2011,SI}. For $m=n=1$, we obtain the microwave intensity shown by the red time trace in Fig.~\ref{fig3:wide}c. Since the SPDC excitation probability in our experiment is far less than unity, this signal is nearly entirely noise added from heating of the hybridized electromechanical modes due to parasitic absorption of pump light. Further, the decay of this added noise exhibits fast and slow components relative to the repetition rate of our experiment. In particular, the slow component is responsible for the non-zero microwave intensity prior to the optical pulse. Detailed measurements of the heating dynamics of the transducer are presented in Ref.~\cite{SI}.

To measure microwave-optical cross-correlations, we perform conditional microwave readout by triggering the heterodyne measurement based on the occurrence of an optical click in an experimental trial. Following the same inversion process used for the unconditional microwave field, we can obtain the moments of the microwave field conditioned on optical detection, $\bar{C}_{mn}|_{\mathrm{click}} = \langle\hat{A}^{\dag} \hat{C}^{\dag m} \hat{C}^n \hat{A} \rangle / \langle \hat{A}^{\dag} \hat{A} \rangle$. Here, $\hat{A}$ refers to the temporal mode defined by gating the optical waveguide mode, $\hat{a}_{\mathrm{out}}$ in the time window indicated by the gray shaded region in Fig.~\ref{fig3:wide}b. The result for $m=n=1$ is shown by the purple time trace in Fig.~\ref{fig3:wide}c. We observe that detection of an optical photon is correlated with substantially higher microwave intensity than that of the unconditional state. The temporal shape of the conditional signal is in good agreement with the result of a numerical simulation of our system \cite{SI}. 

Dividing the conditional and unconditional microwave intensity traces recorded in Fig.~\ref{fig3:wide}c, we obtain the normalized microwave-optical intensity cross-correlation function, 
\begin{equation}
    g^{(2)}_{AC}(\tau) = \frac{\langle\hat{A}^{\dag} \hat{C}^{\dag}(\tau) \hat{C}(\tau) \hat{A} \rangle} {\langle \hat{A}^{\dag} \hat{A} \rangle \langle \hat{C}^{\dag}(\tau) \hat{C}(\tau) \rangle}
    \label{eq: g2_ac}
\end{equation}
In Fig.~\ref{fig3:wide}d, we plot this function sampled at three representative time delays as indicated by vertical dashed lines in Fig.~\ref{fig3:wide}c. $\tau_\mathrm{o}$ is the delay corresponding to the maximum conditional microwave intensity, $\bar{C}_{11}|_{\mathrm{click}}$, and $\tau_\pm = \tau_\mathrm{o} \pm 800$~ns are offset from $\tau_\mathrm{o}$ in opposite directions by five times the FWHM duration of the optical pump pulse. We measure $g^{(2)}_{AC}(\tau_o) = 3.90^{+0.093}_{-0.093}$, indicating strongly correlated microwave and optical emission at this time delay. The error bars for this observation and subsequent correlation functions referred to in the text are determined via a bootstrapping procedure over the dataset of heterodyne voltage samples \cite{SI}, and represent a confidence interval spanning two standard deviations about the mean. We observe that the microwave-optical correlations disappear at times well before and after the optical pump pulse as evinced by the near-unity values of $g^{(2)}_{AC}(\tau_+)=1.00^{+0.08}_{-0.08}$ and $g^{(2)}_{AC}(\tau_-)=0.94^{+0.27}_{-0.27}$. 

For classical microwave-optical states, $g^{(2)}_{AC}$ is bounded by a Cauchy-Schwarz inequality, $g^{(2)}_{AC} \leq \sqrt{{g^{(2)}_{AA}}{g^{(2)}_{CC}}}$ \cite{clauser1974experimental, Kuzmich2003, Riedinger2016}. Here, $g^{(2)}_{AA}$ and $ g^{(2)}_{CC}$ are the normalized intensity autocorrelation functions of the unconditional optical and microwave temporal modes, $\hat{A}$ and $ \hat{C}$ respectively, and are defined in a manner similar to Eq.~(\ref{eq: g2_ac}). Using the moment inversion procedure with the unconditional microwave voltage samples, we have measured $g^{(2)}_{CC}=\bar{C}_{22}/(\bar{C}_{11})^2=1.91^{+0.14}_{-0.14}$ at $\tau=\tau_\mathrm{o}$. This is consistent with the theoretically expected value of 2 for a thermal state. An explicit measurement of $g^{(2)}_{AA}$ with our current device is impractical given the low coincidence rate expected in a Hanbury-Brown-Twiss measurement. In principle, since optomechanical scattering from an acoustic mode in a thermal state is expected to produce $g^{(2)}_{AA}=2$, we expect the classical upper bound, $g^{(2)}_{AC} \leq 2$. Our observation that $g^{(2)}_{AC}(\tau_o)$ exceeds this classical bound for thermal states by over twenty standard deviations serves as a promising signature of non-classical statistics of the microwave-optical states. Further, by performing this cross-correlation experiment with increasing pump power, which is accompanied by increasing thermal noise in the modes, we observe that $g^{(2)}_{AC} (\tau_\mathrm{o})$ monotonically approaches the value of 2 \cite{SI}.

To unambiguously verify the non-classical nature of the microwave-optical photon pairs, we measure the normalized second order intensity correlation function of the microwave state conditioned on an optical click, ${g^{(2)}_{CC}|_{\mathrm{click}} = \bar{C}_{22}|_{\mathrm{click}}/(\bar{C}_{11}|_{\mathrm{click}})^2}$. For noiseless SPDC in the weak pump regime, we expect $g^{(2)}_{CC}|_{\mathrm{click}}=0$ with the detection of an optical photon heralding a pure single photon in the microwave mode. In practice, the value of $g^{(2)}_{CC}|_{\mathrm{click}}$ will be higher due to noise added during transduction. Classical microwave-optical states are expected to satisfy the inequality, $g^{(2)}_{CC}|_{\mathrm{click}} \geq 1$ \cite{SI}. Violation of this inequality signifies the observation of photon anti-bunching in the conditionally prepared microwave state. With the conditional heterodyne voltage samples collected in our experiment, we observe $g^{(2)}_{CC}(\tau_\mathrm{o})|_{\mathrm{click}} = 0.42^{+0.27}_{-0.28}$. As shown in Fig.~\ref{fig3:wide}e, this observation is below the classical bound of unity by 2.1 standard deviations. This corresponds to a probability of 1.7\% for the null hypothesis of conditional preparation of a classical microwave state. Details of the data analysis and the underlying probability distribution of $g^{(2)}_{CC}(\tau_o)|_{\mathrm{click}}$ are provided in the supplementary information \cite{SI}. We emphasize that all the normalized correlation functions referred to in the text are, by definition, independent of the gain, $G$ of the amplification chain in the microwave heterodyne measurement. As a result, even though the values of the moments, $\bar{C}_{mn}$ and ${\bar{C}_{mn}|_{\mathrm{click}}}$ depend on the absolute accuracy of the calibrated gain, our inferences of non-classical statistics are based on normalized correlation functions, which do not require this calibration.  

The non-classical photon pairs generated from our transducer can be used to entangle distant quantum circuits by adopting the DLCZ protocol \cite{Duan2001}. Following this scheme, interference of optical photons from two distant transducers and subsequent single photon detection can be used to herald entanglement between remote microwave nodes \cite{Krastanov2021}. In this work, by conditionally preparing non-classical microwave states using detection of optical photons, we demonstrate both a capable transducer device and key experimental techniques for such a remote entanglement scheme. Towards entanglement of remote quantum processors, the microwave emission from our transducer chip can be routed and efficiently absorbed in a superconducting qubit \cite{Kurpiers2018} residing in a separate processor module \cite{JILA_Nondestructive}. The use of a stronger piezoelectric material such as lithium niobate in our transducer \cite{Jiang2022opticallyheralded,  Weaver2022} would improve the electromechanical conversion efficiency, and thereby, mitigate the primary microwave loss channel in our experiment. Transducer heating due to optical absorption can be reduced through better thermalization of the acoustic mode with the substrate \cite{2DOMC}, thereby improving the fidelity of the photon pair for a given emission rate. Finally, the strongly hybridized electromechanical modes in our transducer provide two frequency bins to encode a qubit in both the optical and microwave emission and can allow the generation of microwave-optical Bell states \cite{Zhong2020FreqBin}. 


\bibliography{refs}

\begin{acknowledgments}
The authors thank M. Mirhosseini, M. Kalaee, A. Sipahigil and J. Banker for contributions in the early stages of this work,  E. Kim, A. Butler, G. Kim, S. Sonar, U. Hatipoglu and J. Rochman for helpful discussions, and B. Baker and M. McCoy for experimental support. We appreciate MIT Lincoln Laboratories for providing the traveling-wave parametric amplifier used in the microwave readout chain in our experimental setup. NbN deposition during the fabrication process was performed at the Jet Propulsion Laboratory. This work was supported by the ARO/LPS Cross Quantum Technology Systems program (grant W911NF-18-1-0103), the U.S. Department of Energy Office of Science National Quantum Information Science Research Centers (Q-NEXT, award DE-AC02-06CH11357), the Institute for Quantum Information and Matter (IQIM), an NSF Physics Frontiers Center (grant PHY-1125565) with support from the Gordon and Betty Moore Foundation, the Kavli Nanoscience Institute at Caltech, and the AWS Center for Quantum Computing. L.J. acknowledges support from the AFRL (FA8649-21-P-0781), NSF (ERC-1941583, OMA-2137642), and the Packard Foundation (2020-71479). S.M. acknowledges support from the IQIM Postdoctoral Fellowship. 
 \\
\end{acknowledgments}




\end{document}


\title{Supplementary Information for ``Non-classical microwave-optical photon pair generation with a chip-scale transducer"}

\author{Srujan Meesala}
\thanks{These authors contributed equally}
\affiliation{Kavli Nanoscience Institute and Thomas J. Watson, Sr., Laboratory of Applied Physics, California Institute of Technology, Pasadena, California 91125, USA}
\affiliation{Institute for Quantum Information and Matter, California Institute of Technology, Pasadena, California 91125, USA}
\author{Steven Wood}
\thanks{These authors contributed equally}
\affiliation{Kavli Nanoscience Institute and Thomas J. Watson, Sr., Laboratory of Applied Physics, California Institute of Technology, Pasadena, California 91125, USA}
\affiliation{Institute for Quantum Information and Matter, California Institute of Technology, Pasadena, California 91125, USA}
\author{David Lake}
\thanks{These authors contributed equally}
\affiliation{Kavli Nanoscience Institute and Thomas J. Watson, Sr., Laboratory of Applied Physics, California Institute of Technology, Pasadena, California 91125, USA}
\affiliation{Institute for Quantum Information and Matter, California Institute of Technology, Pasadena, California 91125, USA}
\author{Piero Chiappina}
\affiliation{Kavli Nanoscience Institute and Thomas J. Watson, Sr., Laboratory of Applied Physics, California Institute of Technology, Pasadena, California 91125, USA}
\affiliation{Institute for Quantum Information and Matter, California Institute of Technology, Pasadena, California 91125, USA}

\author{Changchun Zhong}
\affiliation{Pritzker School of Molecular Engineering, The University of Chicago, Chicago, IL 60637, USA}

\author{Andrew D. Beyer}
\affiliation{Jet Propulsion Laboratory, California Institute of Technology, 4800 Oak Grove Dr, Pasadena, California 91109, USA}

\author{Matthew D. Shaw}
\affiliation{Jet Propulsion Laboratory, California Institute of Technology, 4800 Oak Grove Dr, Pasadena, California 91109, USA}

\author{Liang Jiang}
\affiliation{Pritzker School of Molecular Engineering, The University of Chicago, Chicago, IL 60637, USA}

\author{Oskar~Painter}
\email{opainter@caltech.edu}
\homepage{http://copilot.caltech.edu}
\affiliation{Kavli Nanoscience Institute and Thomas J. Watson, Sr., Laboratory of Applied Physics, California Institute of Technology, Pasadena, California 91125, USA}
\affiliation{Institute for Quantum Information and Matter, California Institute of Technology, Pasadena, California 91125, USA}
\affiliation{Center for Quantum Computing, Amazon Web Services, Pasadena, California 91125, USA}

\date{\today}

\maketitle

\section{Notation}

\begin{table}[ht]
    \centering
    \begin{tabular}{p{0.35\linewidth} | p{0.6\linewidth}}
      Symbol  & Meaning \\ \hline
      $\hat{a}, \hat{b}, \hat{c}$ & operators for optical, acoustic, and microwave modes of the transducer \newline \\
      $\hat{c}_+, \hat{c}_-$ & operators for hybridized electromechanical modes \newline  \\
      $\hat{a}_{\mathrm{in}}, \hat{c}_{\mathrm{in}}, (\hat{a}_{\mathrm{out}}, \hat{c}_{\mathrm{out}})$ & operators for optical and microwave input (output) modes in coupling waveguides \newline \\
      $\hat{A},\hat{C}$ & operators for optical and microwave temporal modes in coupling waveguides  \newline \\
      $\bar{C}_{mn}$ & moments of the temporal microwave mode \newline  \\ 
      $\bar{C}_{mn}|_{\mathrm{click}}$ & moments of the temporal microwave mode conditioned on an optical click \\ 
    \end{tabular}
    \caption{Summary of the various modes and related quantities defined in the main text.}
    \label{tab:symbols}
\end{table}


\section{Device fabrication process}
\begin{figure*}
\includegraphics[width=\textwidth,scale=1]{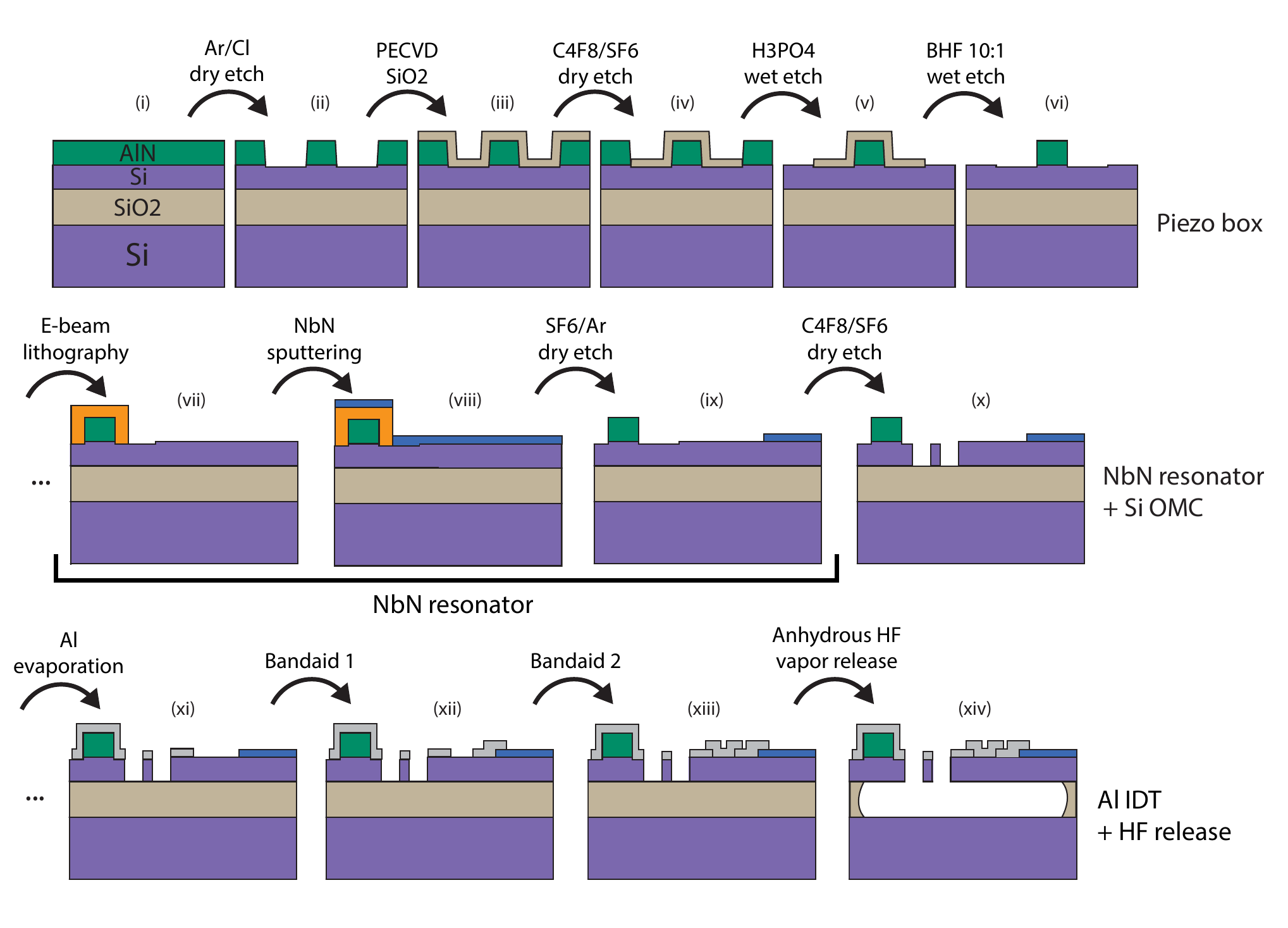}
\caption{Device fabrication process. Images are not to scale. \textbf{(i)} Sputter deposition of 300~nm thick c-axis AlN piezoelectric film (grown by OEM group; stress $T = +55$~MPa; (002) XRD peak of full-width at half-maximum = 1.79°) on a silicon on insulator substrate (Si device layer [float zone grown, 220~nm thick, $\rho \geq 5 \text{k} \Omega \text{-cm}$]; buried oxide layer [3 \textmu m thick, silicon dioxide]; Si handle [Czochralski grown, 750~µm thick, $\rho \geq 5 \text{k} \Omega \text{-cm}$]). \textbf{(ii)} AlN trench etch with Ar/Cl$_{2}$ chemistry to define the perimeter of the piezo resonator via a small trench of width $\sim$100~nm. \textbf{(iii)} Conformal deposition of a SiO$_{x}$ hard mask via plasma enhanced chemical vapor deposition. \textbf{(iv)} Patterning of the SiO$_{x}$ mask via dry etching with C$_{4}$F$_{8}$/SF$_{6}$ chemistry. \textbf{(v)} Removal of the remaining AlN on the chip with a H$_{3}$PO$_{4}$ wet etch. \textbf{(vi)} SiO$_{x}$ mask removal with 10:1 buffered oxide etchant. \textbf{(vii)} Protection of the piezo-acoustic and OMC regions with a resist mask prior to deposition of NbN. \textbf{(viii)} Deposition of a 10 nm thick film of NbN via an RF sputtetring process. \textbf{(ix)} Dry etch of NbN with SF$_{6}$/Ar chemistry to define the high-impedance microwave resonator. \textbf{(x)} Dry etch of silicon with C$_{4}$F$_{8}$/SF$_{6}$ chemistry to define OMC, acoustic shield and optical waveguide. \textbf{(xi)} Deposition of Al electrodes for the piezo-acoustic resonator using angled electron-beam evaporation. \textbf{(xii, xiii)} Bandage steps with Ar milling followed by Al evaporation. \textbf{(xiv)} Etching of the buried oxide (BOX) layer using anhydrous vapor HF to release the device membrane.
}
\label{si:fabrication_fig:wide}
\end{figure*}

The fabrication process for the transducer chip is illustrated in Fig.~\ref{si:fabrication_fig:wide} and described in the caption. Masks for all steps are patterned in ZEP-520A resist via electron-beam lithography on a Raith EBPG 5200 tool. All dry etching is performed in Oxford Plasmalab 100 inductive coupled plasma reactive ion etching (ICP RIE) tools. The process flow can be sub-divided into sections used to define various portions of the transducer device. Steps (i)-(vi) complete the definition of the AlN box essential for the piezoacoustic cavity. The combination of dry and wet etch steps ensures that the dimensions of the AlN box are precisely defined while the silicon device layer is undamaged on most of the chip. This is important to achieve optical, mechanical, and microwave modes with high quality factors. Steps (vii)-(ix) define the NbN resonator and step (x) defines the OMC. Steps (xi)-(xiii) are use to define aluminum electrodes on the piezo-resonator and galvanically connect them to the NbN resonator using bandage steps. The bandaid steps involve in situ Ar milling for two minute and six minute durations respectively to clear the surface of NbN and Al prior to Al bandaid evaporation. To provide optical fiber access to coupler sections at the end of the silicon photonic waveguides, we clear a portion of the SOI substrate up to a depth of 150~$\mathrm{\mu}$m using a deep reactive ion etch at the edge of the chip. Finally, the buried oxide (BOX) layer is etched in anhydrous vapor HF to release the device membrane.

\section{Microwave circuit design}

The kinetic inductance resonator used in our transducer is fabricated from an NbN film of 10~nm thickness and 50~pH/sq sheet inductance. The meandering ladder geometry described in the main text is comprised of  2~$\mathrm{\mu m} \times 1~\mathrm{\mu m}$ rectangular loops formed from traces of width 130~nm. We use extended electrical terminals of length 200~$\mathrm{\mu m}$ and width 1$~\mathrm{\mu m}$ to spatially separate the high kinetic inductance section from the OMC. The resonator is designed to achieve a target frequency of 5.0~GHz for the fundamental mode with a capacitance, $C_{\mathrm{res}}=7.1$~fF, which includes a small contribution of 0.27~fF from the electrodes on the piezo-acoustic resonator. Nearly the entire inductance of the resonator mode is due to the kinetic inductance of the superconducting film. The use of closed superconducting loops in the resonator allows for tuning of the kinetic inductance, $L_k$, via a DC supercurrent, $I$, induced by an external magnetic field according to the relation $L_{k} \approx L_{k,0}[1 + (I/I_{*})^2]$ as shown in \cite{Xu2019}. Here $L_{k,0}$ is the kinetic inductance at zero magnetic field. $I_{*} \gg I$ is a characteristic current on the order of the critical current of the nanowire \cite{Zmuidzinas2012}. This relation leads to quadratic tuning of the resonator frequency in response to an external magnetic field as observed in Fig.~2b of the main text. The resonator is capacitively coupled to an on-chip 50~$\mathrm{\Omega}$ coplanar waveguide (CPW) patterned in NbN to achieve an external coupling rate, $\kappa_{e,c}/2\pi=1.3$~MHz. On the chip used for the experiments in this work, sixteen transducer devices were laid out in groups of four with each group addressed by a separate CPW. Adjacent resonators on the same CPW were designed to be detuned from each other by a frequency spacing of 100~MHz, much larger than the cross-coupling, which was estimated to be less than 1~MHz from simulation. This frequency multiplexing approach allows us to increase the number of transducers available for testing on the chip while ensuring minimal microwave crosstalk.

\section{Piezo-optomechanics design}
\begin{figure*}
\includegraphics[width=\textwidth,scale=1]{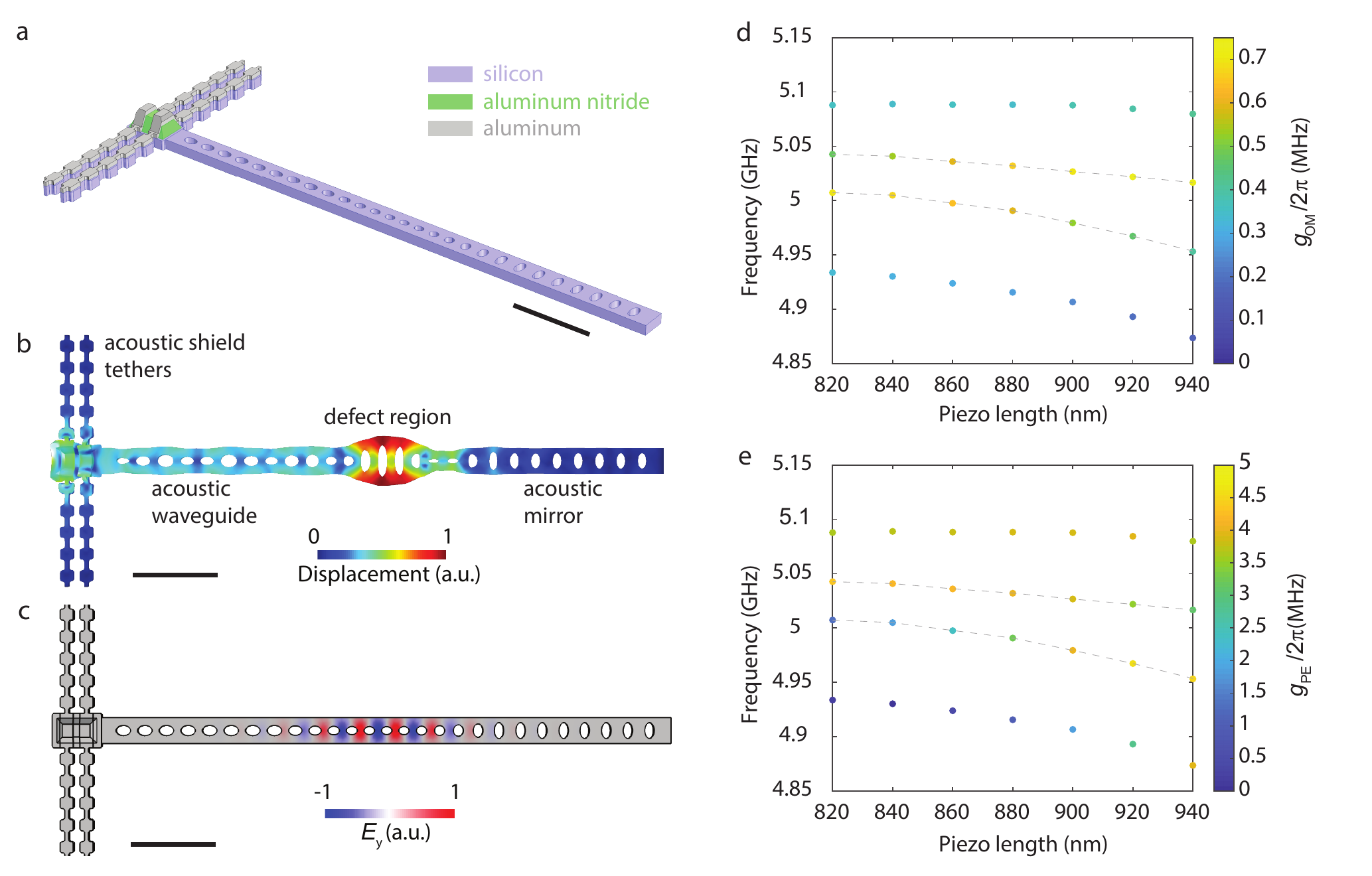}
\caption{Piezo-optomechanics design. \textbf{a.} Geometry and material composition of the piezo-optomechanical transducer. The black scale bar corresponds to 2 \textmu m \textbf{b,c.} Top-down views detailing the displacement profile of a hybridized acoustic mode and the transverse electric field of the fundamental optical mode of the transducer respectively. \textbf{d,e.} Simulated optomechanical coupling, $g_\text{om}$, and piezoelectric coupling, $g_\text{pe}$, of the acoustic modes of the transducer as a function of length of the piezo-acoustic cavity. Dashed lines in both panels are shown as a guide to follow the modes with maximum $g_\text{om}$.}
\label{si:design_fig}
\end{figure*}

The piezo-optomechanical transducer in our work is realized by attaching a half-wavelength AlN piezo-acoustic cavity to a silicon OMC as shown in Fig.~\ref{si:design_fig}a. Figure~\ref{si:design_fig}b,c shows the optical and acoustic modes used for transduction. The energy of the acoustic mode is split between the piezo-acoustic and OMC regions, which are designed independently to support acoustic resonances at 5~GHz with large piezo-electric and optomechanical coupling respectively. Similar to the device demonstrated in \cite{Mirhosseini2020}, hybridization of acoustic modes in these two portions is achieved through the connecting OMC section, whose band structure is designed to provide a bandgap for optical photons and waveguiding for acoustic phonons at the frequencies of interest. The primary difference compared to the previous design is the change to an ultra-low mode volume piezo-acoustic cavity to reduce energy participation in the piezo region, which is lossier for acoustics when compared to silicon. This design change is possible without sacrificing piezoelectric coupling since the impedance of the kinetic inductance resonator is roughly an order of magnitude larger than that of the transmon qubit used in \cite{Mirhosseini2020}. 

Figures~\ref{si:design_fig}d,e show the hybridization of the piezo-acoustic and OMC modes via a sweep of the length of the piezo box. We observe strong hybridization of two distinct branches that are piezo-like and OMC-like as highlighted by the dashed lines with maximal hybridization occurring at a piezo length of 880 nm. On the transducer chip used in our experiments, we fabricate devices with piezo length swept over a range of 20 nm about this nominal value to compensate for fabrication disorder. Clamping loss of the piezo-acoustic resonator is minimized by using tethers with modulated width as shown in Figs.~\ref{si:design_fig}a,b. The tethers are designed to support a 1.5 GHz wide acoustic bandgap centered around the transducer acoustic modes. 

The OMC is evanescently coupled to a suspended silicon waveguide, which terminates in a coupler section at the edge of the chip. We use a millimeter length scale waveguide to distance the NbN circuit from scattering of pump light generated near the optical coupler section. The waveguide is designed with alternating curved sections in order to increase robustness against buckling from intrinsic stress in the silicon device layer. Adiabatic tethers are used to anchor the waveguide in order to reduce scattering loss and maintain high optical collection efficiency.

\section{Measurement Setup}

\begin{figure*}
\includegraphics[width=\textwidth,scale=1]{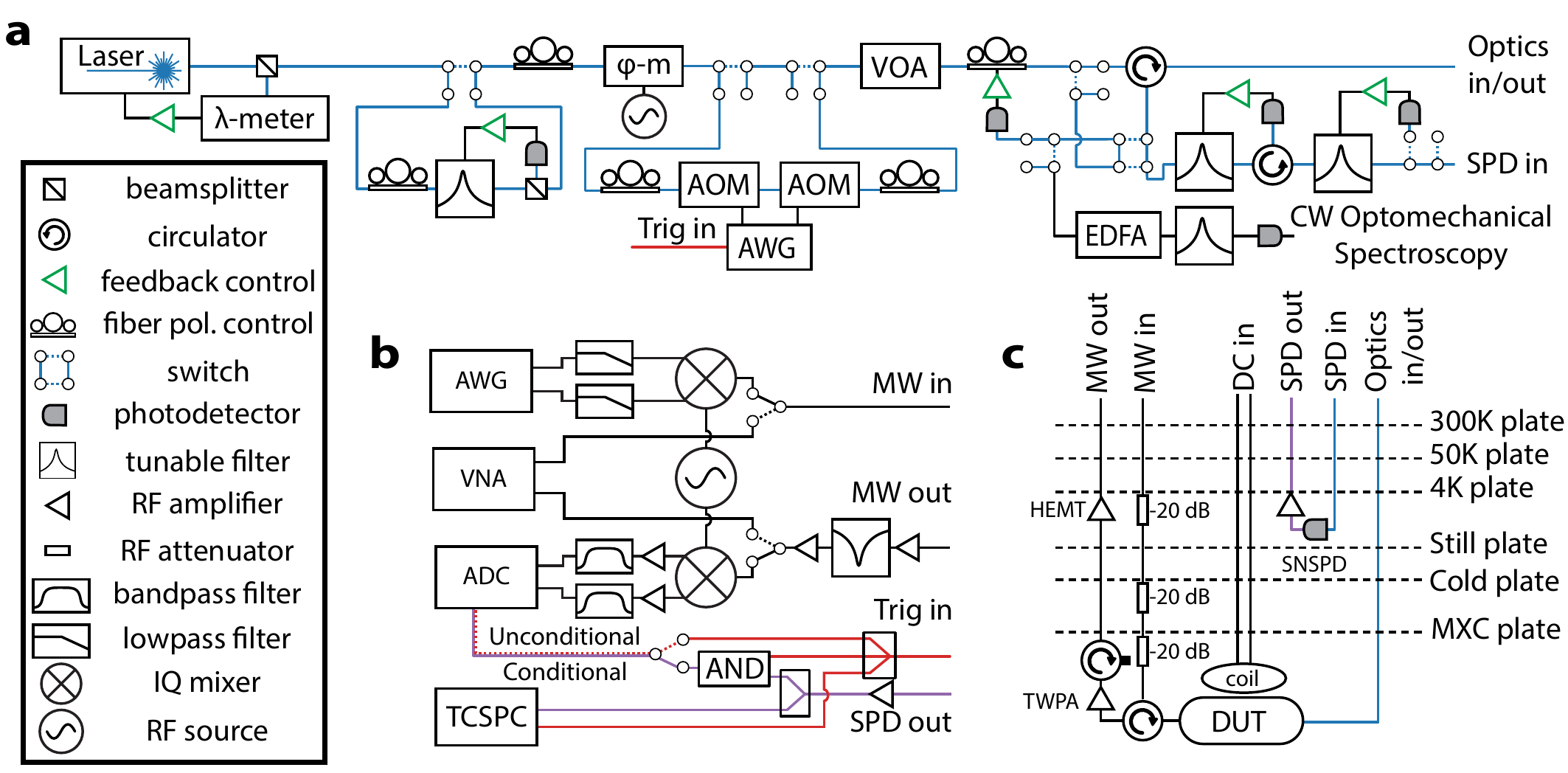}
\caption{Experimental setup. \textbf{a.} Optical pulse synthesis and detection. An external cavity diode laser is wavelength locked to a wavemeter, and a 50 MHz tunable narrowband filter is used to remove spontaneous emission noise. Optical pulse shaping is achieved by modulating the laser with two acousto-optic modulators (AOMs) connected in series. The envelope for the optical pulses is generated by an arbitrary waveform generator (AWG). A variable optical attenuator (VOA) is used to adjust the optical power level sent to the device. Excitation of the device is performed via an optical circulator through which the device reflection and emission can be routed to two separate paths to perform either continuous wave (CW) spectroscopy or pulsed single photon counting. For CW spectroscopy, the reflected optical pump along with the optomechanically transduced signal is amplified with an erbium doped fiber amplifier (EDFA) and detected on a high-speed photodetector. On the single photon counting path, pump light is filtered using two tunable Fabry–Perot filter cavities (Stable Laser Systems) before directing the optical signal to the single photon detector (SPD) in the dilution fridge. \textbf{b.} Microwave pulse synthesis and detection. For measurements that require electrical excitation of the transducer, pulsed input signals to the fridge are generated at the intermediate frequency (IF) by an arbitrary waveform generator (AWG) and subsequently upconverted to $\sim$5GHz on an IQ mixer. Likewise, for CW microwave spectroscopy, a vector network analyzer (VNA) may be switched into the setup. The microwave output from the fridge passes though a tunable bandstop filter to remove the TWPA pump tone, and is down-converted, amplified and filtered before acquisition on an analog to digital converter (ADC). For unconditional microwave readout, the master delay generator for the experiment generates the trigger signal for the ADC. For microwave readout conditioned on an optical click generated from SPDC, the trigger signal for the ADC is generated by performing a logical AND operation between optical click events from the SPD and a trigger bit from the master delay generator spanning the gating window for collection of the optical signal. \textbf{c.} Dilution refrigerator configuration. The microwave (MW) input line contains attenuators on the 4K, cold, and mixing chamber plates for thermalization of the coaxial cables. The MW output line passes through an amplification chain consisting of a Traveling-Wave Parametric Amplifier (TWPA) mounted on the mixing plate and a High Electron Mobility Transistor (HEMT) amplifier mounted on the 4K plate. The TWPA is driven by a strong pump tone sent along a separate line and inserted via a directional coupler placed before the amplifier input (not shown). The superconducting nanowire single photon detector (SNSPD) is mounted to the still plate and the electrical output signal is amplified on the 4K plate before being sent to room temperature electronics.}
\label{si:measurement_fig}
\end{figure*}

The measurement setup used in this work is detailed in Fig.~\ref{si:measurement_fig}. For SPDC experiments, trigger signals from a master digital delay generator are used to synchronize optical pump pulses with the timing window used for optical and microwave readout. Optical pump pulses with $>120$ dB extinction are generated using two analog acousto-optic modulators (G\&H Photonics) and are routed via a circulator into the `Optics in/out' path towards the device in the dilution fridge as shown in Fig.\ref{si:measurement_fig}a. The optical emission from the device passes through the same circulator and is directed to a pump filtering setup prior to single photon detection (SPD) along the `SPD in' path in Fig.~\ref{si:measurement_fig}a. The filtering setup comprises  two tunable Fabry–Perot filter cavities (Stable Laser Systems) in series and provides 104~dB extinction for a pump detuning of 5~GHz along with a  transmission bandwidth of 2.7~MHz. This transmission bandwidth naturally excludes emission due to optomechanical scattering by other transducer modes besides the two hybridized electromechanical modes of interest, $\hat{c}_\pm$. During the experiment, transmission through the filters is checked every four minutes and a lock sequence is initiated if the transmission drops below a set threshold. At the beginning of the transmission check, the optical path is set to bypass the fridge via a MEMS switch. An electro-optic phase modulator ($\mathrm{\phi}$-m in Fig.~\ref{si:measurement_fig}a) is used to generate a sideband on the pump tone at the target frequency to which the filters are locked. In our SPDC experiments, this target frequency corresponds to a pump detuning, $\Delta_{\mathrm{a}} = (\omega_+ + \omega_-)/2$. If required, the locking algorithm adjusts the filter cavities to maximize transmission at the target frequency by monitoring the output of each cavity on a separate photodetector. Additionally, during long measurements, we periodically monitor the polarization of the pump light sent to the device and compensate for long term polarization drifts along the excitation path. Active polarization control is preformed by using an electronic polarization controller (Phoenix Photonics) and maximizing the optical pump power reflected by the device.

Optical photon counting is achieved using a tungsten silicide (WSi) superconducting nanowire single photon detector (SNSPD) mounted on the still plate of the dilution refrigerator (BlueFors LD-250) maintained at a temperature of 770~mK. As shown in Fig.~\ref{si:measurement_fig}b, electrical pulses from the SNSPD generated by single optical photon detection events are amplified and split into two separate paths to record the detection time on a time correlated single photon counter (TCSPC) and to trigger microwave readout conditioned on an optical click. For conditional microwave readout, we first generate a logical bit for every optical click by performing a logic level translation of the SNSPD electrical pulse to a TTL signal. We then perform a logical AND operation between this `optical click bit' and a trigger bit provided by the master delay generator defining the duration of the optical gating window. On the other hand, unconditional microwave readout is triggered directly by the master delay generator. An RF switch (MiniCircuits ZASWA-2-50DRA+) is used to switch between unconditional and conditional readout of the microwave output signal. 

The microwave output chain shown along the `MW out' path in Fig.~\ref{si:measurement_fig}c begins with a Josephson traveling-wave parametric amplifier (TWPA, MIT Lincoln Labs) \cite{Macklin2015twpa} mounted on the mixing plate as the first amplification stage. For TWPA operation, we use a CW pump tone at a frequency of $\sim$6.07~GHz added to the amplifier input using a 20~dB directional coupler (not shown in the figure). A dual junction circulator with 40 dB isolation is placed between the directional coupler and the sample to shield the transducer from back-reflected pump. The TWPA is followed by a high mobility electron transistor (HEMT, Low Noise Factory LNF-LNC4\_8C) amplifier mounted on the 4K plate. In the setup outside the fridge as shown in Fig.~\ref{si:measurement_fig}b, we perform additional amplification of the signal and filtering of the TWPA pump tone using a tunable notch filter (Micro Lambda Wireless MLBFR-0212). The microwave signal is then down-converted to an intermediate frequency (IF) of $\sim$100~MHz after mixing with a local oscillator on an IQ mixer (Marki IQ-4509). The I and Q outputs of the mixer are subsequently bandpass-filtered, amplified and recorded independently on two channels of an analog to digital converter (ADC, Alazartech ATS 9360). These two digitized voltage signals correspond to measurement of the real and imaginary quadratures of the output mode of the amplification chain, $\hat{s}_{\mathrm{out}}$. Based on the calibrated gain of our amplification chain, we measured that our heterodyne detection setup has an added noise of $\sim$2.5~quanta referred to the output of the transducer at a signal frequency of $\sim$5.0~GHz. This near-quantum-limited  heterodyne readout enabled by the TWPA is a key enabler for measurements of microwave-optical correlations from our transducer on a reasonable experimental timescale. The setup shown in Fig.~\ref{si:measurement_fig}b additionally allows for pulsed microwave excitation of the transducer as well as spectroscopy with a vector network analyzer (VNA). 

The transducer sample is wire-bonded to a printed circuit board (PCB) with coaxial connectors and is housed in an oxygen-free high thermal conductivity (OFHC) copper package. The microwave resonance frequency of the NbN resonators can be tuned by an external magnetic field generated from an Nb-Ti coil mounted over the sample package. Individual devices on the chip are optically addressed using a lensed fiber (OZ Optics) affixed to the top of a three-axis piezo stepper stack (Attocube Systems) placed in line with the on-chip tapered optical couplers. The entire assembly is enclosed in a cylindrical magnetic shield and is mounted to the mixing plate of the dilution refrigerator cooled to a base temperature $T_\text{f}\approx 20~\text{mK}$.

%
\section{Optical heralding rate contributions}\label{si:heralding_rate_contributions}

\begin{figure}
\includegraphics[width=\columnwidth,scale=1]{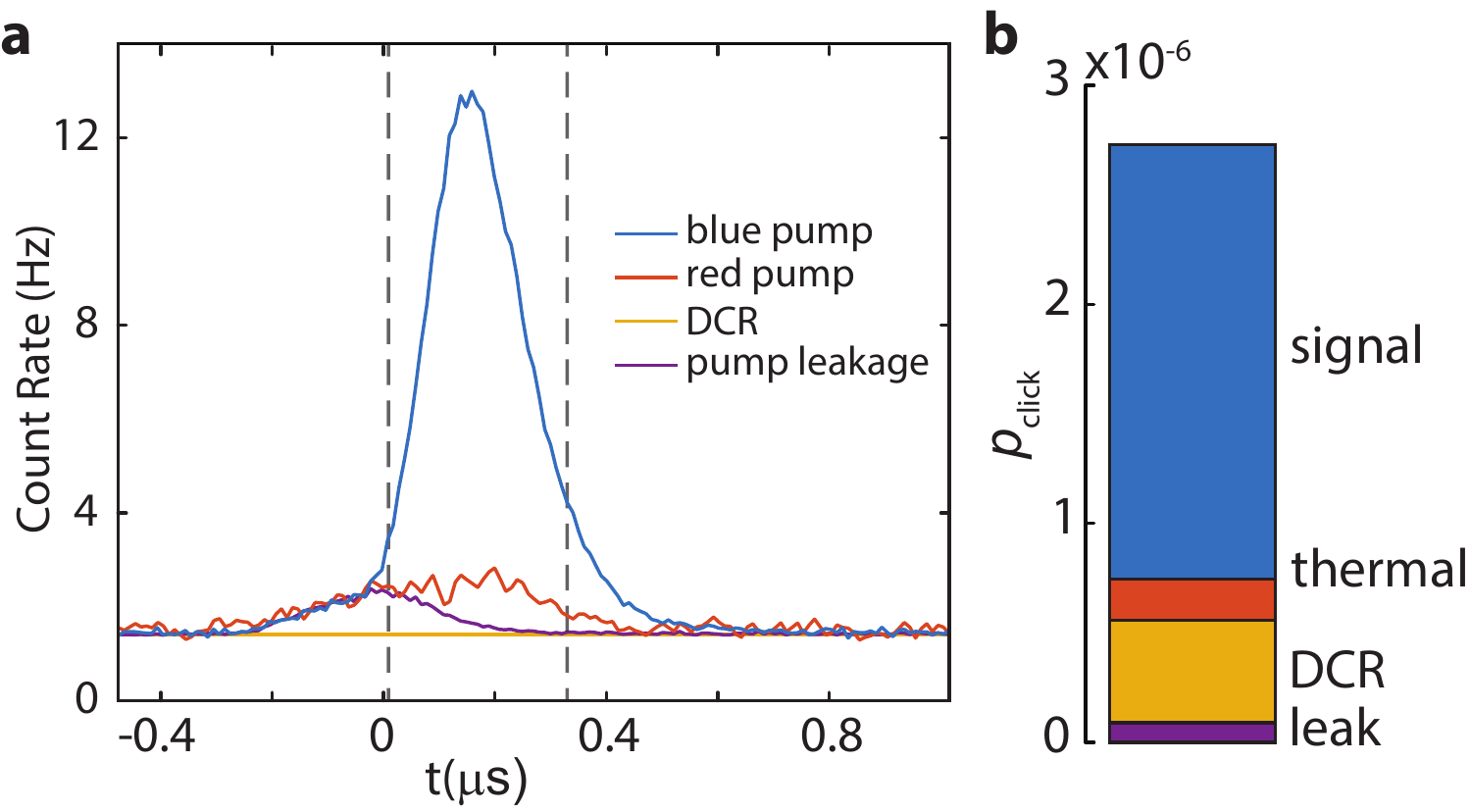}
\caption{\textbf{a.} Optical photon count rate vs time as measured with blue detuned pump pulses, red detuned pump pulses and in calibration measurements for the dark count rate (DCR) and pump leakage through the filter setup. \textbf{b.} Contributions to the optical click probability during the gating window extracted from the data in \textbf{a}.}
\label{si:opticalHR_fig}
\end{figure}

The optical heralding events in our SPDC experiments have a finite noise contribution from pump-induced heating of the hybridized electromechanical modes as well as technical sources in our experimental setup, namely detector dark counts and pump laser leakage. In Fig.~\ref{si:opticalHR_fig}a, we plot the optical photon count rate due to these noise sources as determined from independent measurements. The blue trace corresponds to the total optical signal as shown in Fig.~3a of the main text. Dashed lines bound the gating duration used to select photon clicks that trigger conditional microwave readout. The dark count rate (DCR) of the SNSPD in our experiment is 1.5~Hz and is limited by black-body radiation at wavelengths outside the telecom band guided through the optical fiber. Stray photon flux from laser leakage through the pump filtering setup as indicated by the purple trace in Fig.~\ref{si:opticalHR_fig} is determined by detuning the filter cavity by 100~MHz from the acoustic mode used in the transduction experiment. To determine the noise contribution from the thermal occupation of the acoustic mode, we perform a measurement of optomechanical sideband asymmetry \cite{SMeenehan}. We excite the transducer with the same pump pulse sequence used in the SPDC experiment but with the pump laser tuned to the red mechanical sideband of the optical cavity ($\Delta_{\mathrm{a}} = -(\omega_+ + \omega_-)/2$). The resultant photon flux from this measurement is shown by the red trace in Fig.~\ref{si:opticalHR_fig}a. Integrating the photon counts from these independent measurements over the gating window, we obtain the contributions of each noise source to the total probability of a heralding event, $p_{\mathrm{click}} = 2.7\times10^{-6}$ as shown in Fig.~\ref{si:opticalHR_fig}b. The corresponding fractional contributions are enumerated in Tab.~\ref{tab:op_clicks}. Further, the difference between the count rate under red and blue detuned laser excitation allows us to infer an average thermal occupation of $0.097 \pm 0.019$ for the acoustic mode over the duration of the optical gating window. 

\begin{table}[]
    \centering
    \begin{tabular}{|c|c|c|}
    \hline
      Source of a photon click  &  Fraction of clicks \\
      \hline
      SPDC signal & 0.727 \\
      \hline
      Thermal & 0.069 \\
      \hline
      DCR & 0.171 \\
      \hline
      Pump leakage & 0.033 \\
      \hline
    \end{tabular}
    \caption{Contributions to the optical photon count rate.}
    \label{tab:op_clicks}
\end{table}

%
\section{Microwave moment inversion}\label{sec: moment inversion}
To measure the statistical moments of the microwave emission from the transducer, we adopt the method originally demonstrated in Ref.~\cite{Eichler2011}, which has been widely applied to perform tomography of non-classical microwave radiation emitted by superconducting qubits \cite{Bozyigit2011, Kannan2020, Ferreira2022}. The microwave signal emitted from the device undergoes a series of amplification steps before a record is captured as a digitized, complex-valued voltage signal on the heterodyne setup. This phase-insensitive amplification chain inevitably adds noise and is modelled as \cite{Caves1982,daSilva2010},
\begin{equation}\label{si:amp_flux_eq_exact}
    \hat{s}_\text{out} = \sqrt{G}\hat{c}_\text{out} + \sqrt{G-1}\hat{h}^\dagger,
\end{equation}
where $\hat{c}_\text{out}$, $\hat{s}_\text{out}$ and $\hat{h}^\dagger$ are the bosonic mode operators corresponding to the transducer output, the heterodyne setup output and added noise respectively, with dimensions of $T^{-1/2}$. In the limit of large amplifier gain ($G\gg1$), this relation may be approximated as
\begin{equation}\label{si:amp_flux_eq}
    \hat{s}_\text{out} \approx \sqrt{G}\left(\hat{c}_\text{out} + \hat{h}^\dagger\right),
\end{equation}
Once recorded, we take the inner product of the heterodyne traces with an emission envelope function, $f$ in software. A particular choice of the envelope function corresponds to a measurement of the quadratures of the temporal mode $\hat{S}(t) := \int  \hat{s}_\text{out}(t+t')f^{*}(t') dt'$. With this choice Eq.~(\ref{si:amp_flux_eq}) becomes,
\begin{equation}
    \hat{S} \approx \sqrt{G}\left(\hat{C} + \hat{H}^\dagger\right),
    \label{si:eq:io}
\end{equation}
where $\hat{C}(t) := \int\hat{c}_\text{out}(t+t')f^{*}(t')dt'  $ and $\hat{H}(t) := \int \hat{h}(t+t')f^{*}(t')dt' $. Note that $\hat{C}$ is a temporal mode in the output waveguide, not the internal microwave mode, $\hat{c}$ of the transducer. Thus from a sequence of $N$ heterodyne measurements, a dataset of complex voltage samples $\{S_1,S_2,...S_N\}$ is generated and the statistical moments are numerically calculated as \cite{daSilva2010, Eichler2011}, 
\begin{equation}
    \bar{S}_{mn} := \langle \hat{S}^{\dagger m} \hat{S}^n  \rangle = \frac{1}{N} \sum_{i=1}^N  (S_i^*)^m (S_i)^n.
\end{equation}

The results in the main text require us to determine the statistical moments of $\hat{C}$, which we organize into a moments matrix with elements $\bar{C}_{mn} = \langle \hat{C}^{\dagger m} \hat{C}^n \rangle$. This is accomplished using additional reference measurements in which the transducer is not optically pumped so that the input to the amplifier is a vacuum state. This reference measurement directly gives the anti-normally ordered moments of the added noise as $\bar{H}_{mn} := \langle \hat{H}^m\hat{H}^{\dagger n} \rangle =\bar{S}_{mn}/G^{\frac{n+m}{2}}$.

We experimentally verified that $\bar{H}_{mn}$ is consistent with a thermal state for $m,n\leq2$. Knowledge of the mean occupancy ($n_{\text{th}, H}$) of the state constitutes full knowledge of the statistics for a thermal state. Accordingly, we extract $n_{\text{th}, H}$ from the reference measurements and calculate $\bar{H}_{mn}$ as \cite{Barnett2018},
\begin{equation}
    \bar{H}_{mn} = 
    \begin{cases} 
    m!(n_{\text{th}, H}+1)^m &m=n, \\
    0 & \text{otherwise}. 
    \end{cases}
\end{equation}

Finally, under the assumption that there are no correlations between $\hat{C}$ and $\hat{H}$, Eq.~(\ref{si:eq:io}) can be used to derive the following relation between the moment matrix elements \cite{Eichler2011},
\begin{equation}\label{si:allmoments_eq}
    \bar{S}_{ij} = \bar{T}_{ijmn}\bar{C}_{mn},
\end{equation}
where
\begin{equation}
\bar{T}_{ijmn}=
   \begin{cases} 
      G^{\frac{i+j}{2}}{i \choose m}{j \choose n} \bar{H}_{i-m,j-n} & m\leq i, n\leq j, \\
      0 & \text{otherwise}. 
   \end{cases}
\end{equation}
By inverting Eq.~(\ref{si:allmoments_eq}), we recover $\bar{C}_{mn}$. 

\section{Microwave emission envelope function}\label{si:MW_EM_Section}
\begin{figure}[!h]
\includegraphics[width=\columnwidth,scale=1]{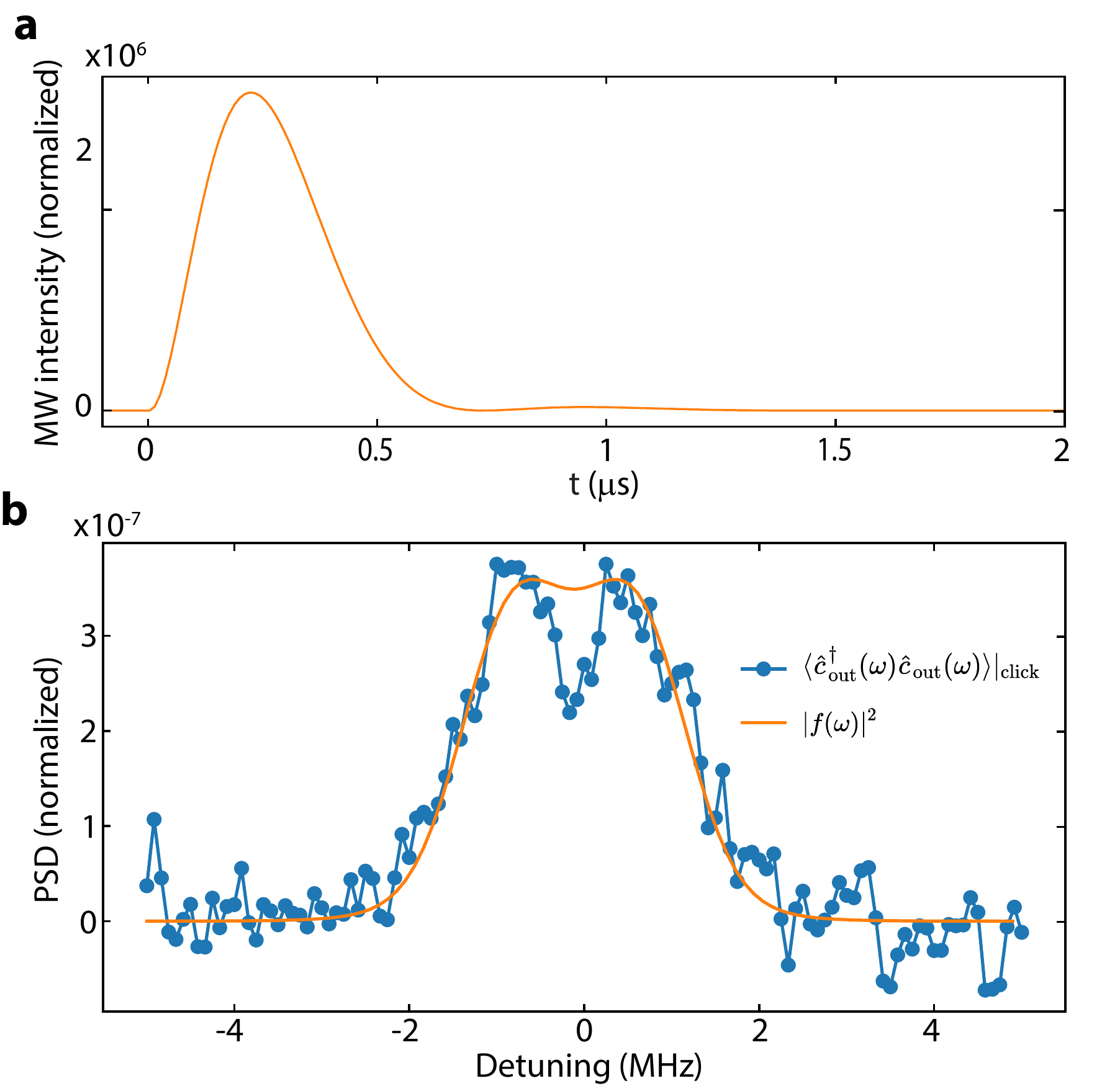}
\caption{\textbf{a.} Simulated temporal profile of microwave emission intensity from the transducer heralded by an optical detection event. \textbf{b.} Comparison of normalized power spectral density (PSD) of the emission envelope function, $f$ (orange trace) and of the experimentally measured transducer microwave output conditioned on an optical click (blue data points). Detuning is defined with respect to the mean frequency of the hybridized electromechanical modes.}
\label{si:evelope_fig}
\end{figure}

We define the two hybridized electro-mechanical modes in our experiment as
\begin{equation}
    \hat{c}_\pm  =  \frac{1}{\sqrt{2}}\left(\hat{b} \pm \hat{c}\right), 
\end{equation}
with center frequencies $\omega_\pm = \omega_b \pm g_\text{pe}$. 

The ideal two-mode squeezed state generated under constant optical pumping on the blue optomechanical sideband ($\Delta = \omega_a - \omega_p = \omega_c$) can be written in the form \cite{Zhong2020FreqBin},
\begin{equation}
    \ket{\psi} = \left[1 + \sqrt{p}\left(\hat{a}_+^{\dag}\hat{c}_+^{\dag} + \hat{a}_-^{\dag}\hat{c}_-^{\dag}\right) + \mathcal{O}(p) \right] \ket{\mathrm{vac}},
\end{equation}
up to a normalization factor. Here $p$ is the probability of an optomechanical scattering event, $\hat{a}_\pm$ are the optical modes which match the frequencies of the hybridized modes $\hat{c}_\pm$ as required by energy conservation. Performing single photon detection on the optical state coupled out of the cavity naturally allows us to discard the vacuum component of $\ket{\psi}$. Further, we can neglect higher order terms, $\mathcal{O}(p)$ in the expansion since $p \ll 1$ for the optical pump power level used in our experiments. In our experiment, the SNSPD has nanosecond timing jitter, which is much smaller than the period of the beat note between the modes $\hat{a}_\pm$ given by $2\pi/2g_{\text{pe}} = 625$~ns. As a result, detection of an SNSPD click in an experimental trial erases frequency information of the optical state. This measurement can be described as a projective measurement onto the state, $ \left(e^{i\phi(t)/2}\hat{a}_+ + e^{-i\phi(t)/2}\hat{a}_-\right)\ket{\text{vac}}$, where the relative phase $\phi(t)=\phi_o+2g_\text{pe}t$ for an optical click received at time $t$. Here $\phi_o$ is the constant relative phase acquired between the two frequency bins along the optical path and $t$ is the emission time of the photon pair defined relative to the beginning of the pump pulse. An optical click at time $t$ can therefore be used to herald the microwave state $\ket{\psi}_\text{click} = \left(e^{-i\phi(t)/2}\hat{c}_+^{\dag} + e^{i\phi(t)/2}\hat{c}_-^{\dag}\right)\ket{\mathrm{vac}}$. 

This picture becomes slightly more complicated when considering the pulsed optical drive used in this work as well as the parameters of the device. The large bandwidth of the pump pulse of 2.75 MHz in comparison to the mode splitting of 1.6 MHz means that the frequency bins cannot be individually well resolved in the optical emission. Further, given the splitting between the frequency bins versus their individual linewidths of 1.0 MHz, our device is not deep in the strong piezoelectric coupling regime. This is evinced by the two peaks with finite overlap in the experimentally measured power spectrum of the conditional microwave emission, $\langle \hat{c}^\dagger_\text{out}(\omega)\hat{c}_\text{out}(\omega) \rangle|_\text{click}$ shown with the blue trace in Fig.~\ref{si:evelope_fig}b. However, the theoretical picture discussed above can guide the parametrization of the envelope function used for microwave readout. Accordingly, we define the microwave emission envelope function as a coherent superposition of two frequency bins centered at $\omega_\pm$,
\begin{equation}
    f(t) = \frac{g(t)}{\sqrt{2}}\left( e^{-i(\omega_+ t+\phi_o/2)} + e^{-i(\omega_- t - \phi_o/2)}\right).
    \label{eq:f_phi}
\end{equation}
The relative phase, $\phi_o$ is assumed to be fixed since the optical pulse duration used in our experiment is short compared to the beat period between the frequency bins. The function, $g(t)$ accounts for the finite linewidth of the frequency bins and is obtained by numerically solving the master equation for our system using the QuTiP software package \cite{Johansson2013}. For this calculation, we use the Hamiltonian in Eq.~(1) of the main text along with experimentally determined coupling and decay rates for the transducer modes. The resulting MW emission intensity is shown in Fig.~\ref{si:evelope_fig}a. $g(t)$ is parameterized as a skewed Gaussian to capture the faster timescale of the rising edge relative to the decay. The corresponding voltage envelope is given by the square root of a skewed Gaussian and is written as
 \begin{equation}
     g(t) = \frac{1}{\sqrt[4]{{2\pi}T_\text{g}^2}}\mathrm{exp}\left(- \frac{t^2}{4T_\text{g}^2} \right)\left[ 1 + \mathrm{erf}\left( \frac{\alpha t}{\sqrt{2}T_\text{g}} \right)\right]^{1/2},
 \end{equation}
up to a normalization factor involving the gain of the microwave amplification chain. Here the Gaussian standard deviation, $T_\text{g}$ = 230~ns and skew factor, $\alpha = 2.0$ are obtained from fitting to the simulated result.  We note that for a microwave photon instantaneously loaded into the resonator, $g(t)$ is expected to follow an exponential decay. However, in our transducer, due to the finite conversion rate of the phonon from SPDC into a microwave photon, $g(t)$ has a finite rise time. As a result, we find that choosing the skewed Gaussian parametrization instead of an exponential decay provides $\sim 4\%$ higher overlap between the emission envelope function and the simulated wavepacket as quantified by their normalized inner product. The power spectrum of the resultant envelope function, $|f(\omega)|^2$ is shown as the orange trace in Fig.~\ref{si:evelope_fig}b. This is observed to be well-matched to the power spectrum of the conditional microwave emission from the transducer obtained from experimentally recorded heterodyne voltage traces. 

\section{Microwave gain calibration}
We calibrate the gain of our microwave heterodyne detection setup by operating the transducer as a frequency converter with the optical pump laser on the red mechanical sideband. The calibration procedure is performed at zero magnetic field, when the acoustic and microwave modes, $\hat{b}, \hat{c}$ are detuned by a frequency separation, $\omega_b - \omega_c = 2\pi\times 12~\mathrm{MHz} \gg 2g_{\mathrm{pe}}$ and are weakly hybridized. An independent measurement of optomechanical sideband asymmetry provides a meter for phonon occupation in the acoustic mode, $\hat{b}$ via the single phonon count rate, $R_o$ \cite{SMeenehan}. We then resonantly excite the acoustic mode via the microwave port and measure the transduced optical photon count rate, $R$. Simultaneously, we record the reflected microwave signal from the transducer, which produces an output power, $P_{\mathrm{det}}$ at the output of the heterodyne setup. The gain, $G$ of the amplification chain can be determined using the series of equations below.

\begin{eqnarray}
    R &=& n_{b,\mathrm{sig}}R_o, \\
    n_{b,\mathrm{sig}}\hbar\omega_b &=& P_{\mathrm{in}}\frac{4\kappa_{e,b}}{\kappa_{i,b}^2}, \\
    P_{\mathrm{det}} &=& G \left( \frac{\kappa_{e,b} - \kappa_{i,b}}{\kappa_{e,b} + \kappa_{i,b}} \right)^2 P_{\mathrm{in}}.
\end{eqnarray}

Here, $n_{b,\text{sig}}$ is the occupation of the acoustic mode due to the microwave drive obtained after subtracting the thermal component of the transduced signal. $P_\text{in}$ is the input microwave power to the transducer. $\kappa_{b,\text{e}}$ is the coupling rate of the acoustic mode to the microwave waveguide due to weak hybridization with the microwave resonator and $\kappa_{b,\text{i}}$ is the intrinsic linewidth of the acoustic mode. Both rates are determined via a VNA measurement of the acoustic mode in the presence of the optical pump pulses. Following this procedure, we estimate gain, $G$ = 103~dB between the microwave output port of the transducer and the output of the heterodyne setup. While knowledge of the gain allows us to estimate the microwave intensity at the transducer output, we note that the values of the normalized correlation functions measured in our experiment are independent of the absolute accuracy of this calibration. 

\section{TWPA optimization}
The operating point of the TWPA was optimized prior to data acquisition. In order to ensure fast acquisition and unbiased statistics, our operational criterion were (1) maximize the gain and (2) ensure linearity in the response. In our optimization procedure, the frequency and power of the pump tone driving the TWPA were swept, and coherent pulses with amplitude matched to the conditional microwave signal in our experiment were sent to the transducer input port. The result of the heterodyne measurement of the reflected signal from the transducer was used to compute $g^{(2)}_{CC}$. To ensure linearity, we chose a region in the TWPA pump frequency and power space where $g^{(2)}_{CC}=1$ to within one standard deviation. The final operating point was then chosen as the one with maximum gain from this cohort. Over the course of the experiment, linearity of the TWPA was periodically verified by measuring $g^{(2)}_{CC}$ for a weak coherent input. In case of a deviation of $g^{(2)}_{CC}$ from unity by one standard deviation, the experiment was halted and the pump optimization routine was re-run.

%
\section{Transducer heating dynamics}
\begin{figure}
\includegraphics[width=\columnwidth,scale=1]{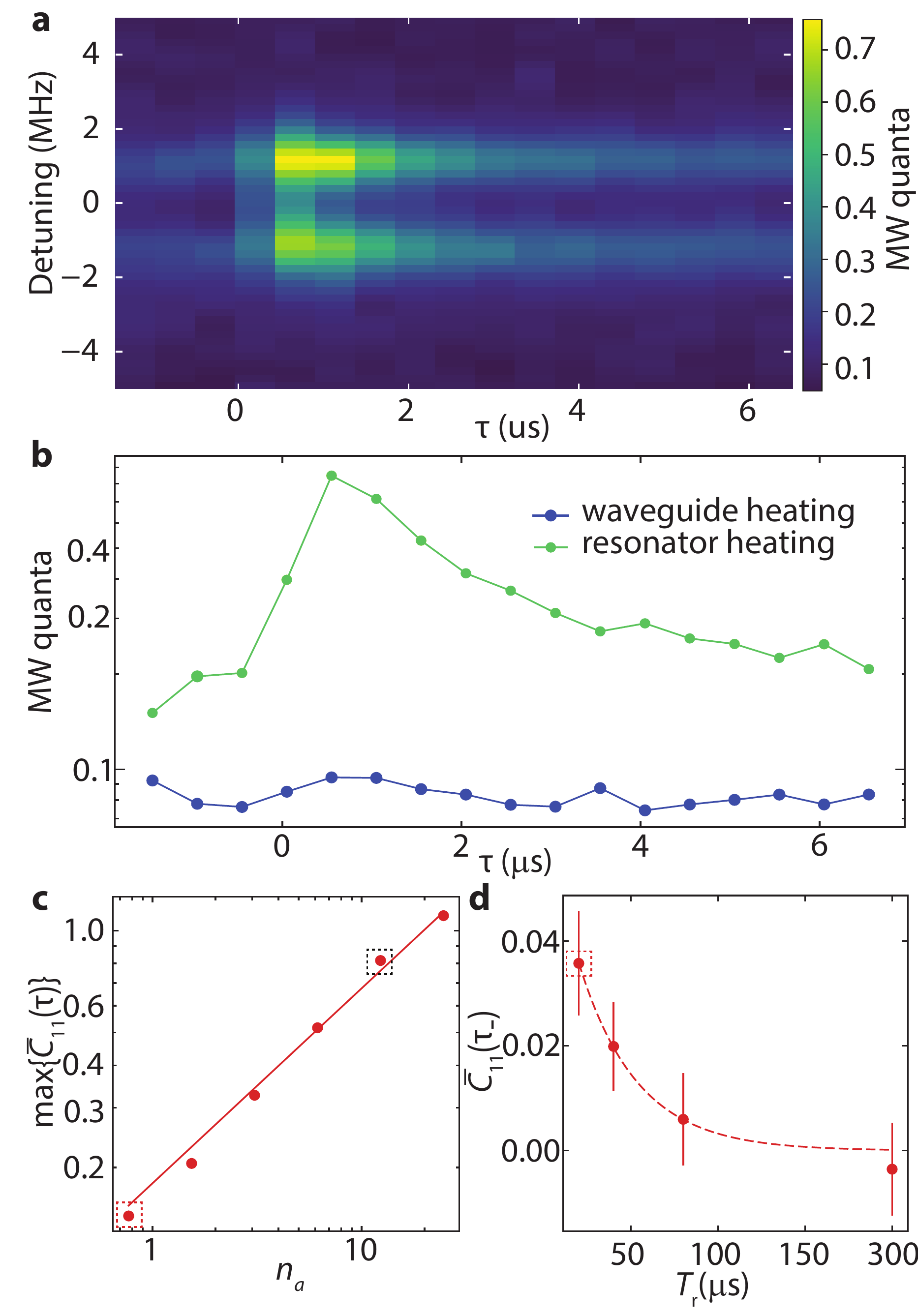}
\caption{Transducer heating dynamics. \textbf{a.} Power spectral density of the unconditional microwave signal as a function of delay, $\tau$ from the center of the optical pump pulse. Detuning is defined with respect to the mean frequency of the hybridized electromechanical modes. Legend indicates microwave quanta emitted at the output port in a 100~kHz bandwidth. Measurement is performed with optical pump pulses of peak power, $n_a$ = 12, pulse duration of 160~ns, and 50~kHz repetition rate. \textbf{b.} Time dependence of waveguide and resonator components of heating extracted by fitting the data in panel \textbf{a.}. \textbf{c.} Thermal emission in the transducer microwave output as a function of optical pump power. Pump power is plotted in terms of peak intra-cavity occupation of the optical mode, $n_a$ and microwave intensity is plotted in terms of occupation of the temporal mode, $\bar{C}_{11}$ measured at the delay, $\tau$ at which it is maximized. All measurements were performed with the 160~ns pulse duration and 50~kHz repetition rate used for data in the main text. Dashed red box denotes the power used for the experiment in the main text ($n_a = 0.8$) while the dashed black box denotes the power used for measurements in panels \textbf{a,b}.  Solid line is a power law fit revealing an exponent, 0.58 $\pm$ 0.03. \textbf{d.} Decay of the slow component of the thermal emission probed by varying the repetition period, $T_r$ of the pump pulse sequence. Microwave quanta in the unconditional state, $\bar{C}_{11}$ are measured versus prior to the pump pulse at the delay $\tau_{-}$ as defined in the main text. Dashed line is an exponential fit revealing a decay time of $33\pm 6~\mathrm{\mu}$s. Dashed red box denotes the repetition rate used for the experiment in the main text.}
\label{si:heating_dynamics}
\end{figure}

To investigate the heating dynamics of the transducer, we performed a series of microwave measurements under pulsed optical excitation. In Fig.~\ref{si:heating_dynamics}a, we show the time-resolved power spectrum of the unconditional microwave emission at the transducer output port. For this measurement, we use the optical pulse sequence ($T_{\mathrm{p}}$ = 160 ns, $50$ kHz repetition rate) used in the SPDC experiments in the main text albeit at a higher pump power, $n_a=12$, which generates higher thermal noise. We observe that majority of the heating occurs in the emission bandwidth of the hybridized electro-mechanical modes. The peak of this thermal emission is noticeably delayed from the optical pump pulse. Additionally, we observe a small but non-zero noise contribution far from the transducer resonances, which we attribute to heating of the microwave waveguide. These resonator and waveguide contributions to the added noise are separated by fitting the power spectrum at each delay to a double Lorentzian function with a constant floor. The fit result is shown in Fig.~\ref{si:heating_dynamics}b, and indicates that resonator heating significantly dominates waveguide heating and has a stronger time dependence. This points to parasitic optical absorption in the intrinsic baths of the acoustic and microwave modes as the dominant source of transducer heating. Additionally, we measured the scaling of the intensity of this thermal emission with optical pump power and observed it to be sub-linear as shown by the results plotted in Fig.~\ref{si:heating_dynamics}c. Finally, from the time dynamics of heating in Fig.~\ref{si:heating_dynamics}a, we observe that while most of the thermal emission decays on the timescale of a few microseconds, a small component persists as steady state heating and is observed prior to the arrival of the optical pump pulse. We attribute this to a slower component of the hot bath generated by optical absorption and characterize it by varying the repetition rate of the optical pulses. The result shown in Fig.~\ref{si:heating_dynamics}d confirms slow decay of this component of heating with an exponential decay time of $33\pm 6 \mathrm{\mu}$s. The measurements described in this section were performed on the same device used for the SPDC experiments in the main text, but after a partial warm up and cooldown to base temperature. After this thermal cycle, we observed frequency shifts of the microwave and acoustic modes by a few MHz and an increase in the normal mode splitting. However, the thermal noise quanta generated by the optical pulse sequence used for the experiments in the main text remained nearly the same.

%
\section{Simulation of the conditional microwave state}
As detailed in Eq.\,1 of the main text, the dynamics of the transducer are governed by the Hamiltonian,
\begin{equation}
    \hat{H}_{abc}(t) = \hat{H}_o + \hat{H}_\text{om}(t) + \hat{H}_\text{pe},
\end{equation}
where $\hat{H}_o$, $\hat{H}_\text{om}(t)$, and $\hat{H}_\text{pe}$ contain the mode frequencies, optomechanical interaction, and piezoelectric interaction terms respectively. These are defined as
\begin{subequations}
\begin{align}
    \hat{H}_o &= -\hbar\Delta_a\hat{a}^\dag\hat{a} + \hbar\omega_b\hat{b}^\dag\hat{b} + \hbar\omega_c\hat{c}^\dag\hat{c},\\
    \hat{H}_{\mathrm{om}}(t) &= \hbar G_{\mathrm{om}}(t)(\hat{a}^\dag\hat{b}^\dag + \hat{a}\hat{b}),\\
    \hat{H}_{\mathrm{pe}} &= \hbar g_{\mathrm{pe}}(\hat{b}^\dag\hat{c} + \hat{b}\hat{c}^\dag).
\end{align}
\end{subequations}
We have explicitly included the time dependence of the optomechanical coupling $G_{\mathrm{om}}(t) = \sqrt{n_a(t)} g_{\mathrm{om}}$ due to the temporal shape of the intra-cavity pump photon number, $n_a(t) = {\kappa_{\text{e},a}}/({\Delta_a^2+\kappa_{a}^2/4}) P_\text{in}(t)$. $P_\text{in}(t)$, the optical pump power at the device follows a Gaussian shape with parameters described in the main text. In our experiment, we use a blue detuned pump with $\Delta_a=\omega_b$. Further, our device is in the sideband resolved regime with $\omega_b \gg \kappa_a$. 

In addition to unitary evolution due to couplings among the internal modes of the transducer, the system is also subject to dissipation and heating due to coupling to the environment. This is captured by the master equation,
\begin{equation}\label{si:eq_full_Lim}
     \dot{\hat{\rho}}(t) = \mathcal{L}(t)\hat{\rho}(t).
\end{equation}
Defining the Lindblad superoperator as $\mathcal{D}(\hat{o}) \hat{\rho} = \hat{o}\hat{\rho}\hat{o}^\dagger - \frac{1}{2} \left\{ \hat{o}^\dagger \hat{o}, \hat{\rho} \right\} $ the action of $\mathcal{L}(t)$ on the density matrix is written as
\begin{equation}
    \mathcal{L}(t)\hat{\rho}(t) = -\frac{i}{\hbar}\left[ \hat{H}_{abc}, \hat{\rho} (t)\right]  + \left(\mathcal{L}_a  + \mathcal{L}_b(t) + \mathcal{L}_c(t)\right) \hat{\rho} (t),
\end{equation}
where the superoperators, $\mathcal{L}_a, \mathcal{L}_b, \mathcal{L}_c$ describe the coupling of the respective transducer modes to the environment according to the relations below.
\begin{subequations}
\begin{align} 
    \mathcal{L}_a\hat{\rho}(t) =&\, \kappa_{a} \mathcal{D}(\hat{a})\hat{\rho}(t), \\
    \mathcal{L}_b(t)\hat{\rho}(t) =&\, n_{\text{th},b}(t) \kappa_{\text{i},b} \mathcal{D}(\hat{b}^\dagger)\hat{\rho}(t) \nonumber\\
    &\, + \left[n_\text{th,b}(t)+1\right]\kappa_{i,b} \mathcal{D}(\hat{b})\hat{\rho}(t), \label{SI:eq_diss_mech}\\
    \mathcal{L}_c(t)\hat{\rho}(t) =&\, n_{\text{th},c}(t)\kappa_{\text{i},c} \mathcal{D}(\hat{c}^\dagger)\hat{\rho}(t) \nonumber\\
    &\, + \left[n_{\text{th},c}(t)+1\right]\kappa_{i,c} \mathcal{D}(\hat{c})\hat{\rho}(t)\nonumber\\
    &\,+n_{\text{th},w}\kappa_{\text{e},c} \mathcal{D}(\hat{c}^\dagger)\hat{\rho}(t) \nonumber\\
    &\, + \left[n_{\text{th},w}+1\right]\kappa_{e,c} \mathcal{D}(\hat{c})\hat{\rho}(t).
\end{align}
\end{subequations}
Here the total dissipation rate of the optical mode, $\kappa_a=\kappa_{e,a}+\kappa_{i,a}$. To capture heating of the transducer modes due to parasitic absorption of the optical pump, we assume that the acoustic and microwave modes are coupled to intrinsic baths with time dependent thermal occupation, $n_{\text{th},b}(t)$ and $n_{\text{th},c}(t)$ respectively. $n_{\text{th},w}$, the thermal occupation of the microwave waveguide is set to a constant value much smaller than $n_{\text{th},b}(t)$ and $n_{\text{th},c}(t)$ based on the measurements detailed in Section 11. We note that while these baths may be significantly more complex and involve several components possessing different, time-dependent coupling strengths, our model is aimed at capturing the total influx of thermal excitations into the transducer modes generated by the optical pulse. With this endeavor in mind, we assume that the acoustic (microwave) mode couples to a single intrinsic bath at a fixed dissipation rate, $\kappa_{i,b}$ ($\kappa_{i,c}$) and ascribe the time-dependence of the heating dynamics entirely to the thermal occupation of the bath. This phenomenological approach to heating induced by optical absorption has previously been used to model photon correlations in optomechanics experiments \cite{Hong2017}. 

While Eq.~(\ref{si:eq_full_Lim}) can be numerically solved in principle, this is computationally intensive owing to the large state space involving the three modes of the transducer. Instead, we take advantage of the fact that $\kappa_{a} \gg G_\text{om}$ to adiabatically eliminate the optical mode \cite{Wilson-Rae2008}. Moving into a frame rotating with the mechanical and microwave modes and defining the optomechanical scattering rate, $\Gamma_\text{om}(t) = 4|G_\text{om}(t)|^2/\kappa_{a}$, we arrive at the simplified master equation,
\begin{equation}\label{SI:eq_ad_elim}
     \dot{\hat{\rho}}_r(t) = \mathcal{L}_r(t)\hat{\rho}_r(t),
\end{equation}
where $\hat{\rho}_r = \mathrm{Tr}_a\{ \hat{\rho}\}$ denotes the reduced density matrix spanning only the acoustic and microwave modes. $\mathcal{L}_r$ is defined by its action on the reduced density matrix,
\begin{align}
    \mathcal{L}_r(t)\hat{\rho}_r(t) =&\, -\frac{i}{\hbar}\left[ \hat{H}_\text{pe}, \hat{\rho_r} (t)\right]   \nonumber\\
    &\,+\left(\mathcal{L}_{\text{om}}(t)  + \mathcal{L}_b(t) + \mathcal{L}_c(t)\right) \hat{\rho}_r (t),
\end{align}
where we have introduced optomechanical scattering as a coupling of the reduced system to an effective bath according to the relation,
\begin{equation}
    \mathcal{L}_{\text{om}}(t) \hat{\rho}_r(t) = \Gamma_\text{om}(t) \mathcal{D}(\hat{b}^\dagger)\hat{\rho}_r(t)\label{SI:eq_optical_loss} .
\end{equation}

The Lindblad superoperator, $\mathcal{D}(\hat{b}^\dagger)$ contains a non-number preserving quantum jump operator which adds a phonon to the system due to optomechanical SPDC. This process is naturally correlated with the creation of an optical photon, which is routed with finite collection efficiency to a single photon detector. Given that optical decay ($\kappa_a/2\pi = 1.3$ GHz) occurs on a much faster timescale than the dynamics of the rest of the system ($\kappa_a \gg \kappa_b, g_\text{pe}, \kappa_c$), optical detection is assumed to be instantaneous. Under this approximation, in the event of a click on the optical detector caused by a quantum jump at time $t_\text{J}$, the resultant conditional state of the reduced system is,
\begin{equation}
    \hat{\rho}_\text{J}(t_\text{J}) = \frac{ \hat{b}^\dagger \hat{\rho}_r(t_\text{J}) \hat{b}}{\mathrm{Tr} \left\{ \hat{b}^\dagger \hat{\rho}_r(t_\text{J}) \hat{b} \right\}}. \label{si:eq_jump}
\end{equation}
This state then evolves according to Eq.~(\ref{SI:eq_ad_elim}). Since we operate in the weak pump regime where the integrated jump probability over the optical pulse duration, $\int \Gamma_\text{om}(t) dt \ll 1$, we neglect two-fold SPDC events.

While the treatment above describes the evolution of the internal modes of the transducer, we experimentally measure the emission in the temporal mode, $\hat{C}$ in the transducer microwave output port. This temporal mode is linearly related to the internal microwave mode of the transducer as $\hat{C}(t) := \int \hat{c}_\text{out}(t+t')f^{*}(t') dt' = \sqrt{\kappa_{e,c}} \int \hat{c}(t+t')f^{*}(t') dt'$. We can thus compute the occupation of the temporal mode as 
\begin{eqnarray}
& & \langle\hat{C}^\dag (t) \hat{C} (t) \rangle \notag \\
& = & \kappa_{e,c} \langle \int\hat{c}^\dag (t+t'')f(t'') dt'' \int\hat{c} (t+t')f^{*}(t') dt' \rangle \notag \\
& = & \kappa_{e,c} \int \langle \hat{c}^\dag (t+t'') \hat{c} (t+t')  \rangle f(t'') f^{*}(t') dt''dt' \notag \\
& = & \kappa_{e,c} \int \langle \hat{c}^\dag (t'+\tau ') \hat{c} (t')  \rangle f(t'+\tau ' - t) f^{*}(t' - t) dt'd\tau ', \notag\\
& & \label{eq:CdagC}
\end{eqnarray}
where the correlator, $\langle \hat{c}^\dag (t'+\tau ') \hat{c}(t') \rangle$ inside the integral on the final line can be numerically computed using the quantum regression theorem \cite{GardinerZoller2004}. Further, to model the conditional state due to a quantum jump at time $t_\text{J}$, this correlator is computed using $\hat{\rho}_\text{J}(t_\text{J})$ in Eq.~(\ref{si:eq_jump}) for times, $t>t_{\mathrm{J}}$. For times $t<t_{\mathrm{J}}$, we use the value for the unconditional state. Performing this calculation for a sequence of jump times, $\{t_{\mathrm{J}}\}$ associated with detector clicks received within the optical gating window, we obtain a sequence of conditional time traces for the occupation of the temporal mode, $\{ \langle\hat{C}^\dag \hat{C} \rangle | _{t_{\mathrm{J}}} \}$. The final conditional trace, $\langle\hat{C}^\dag\hat{C} \rangle | _{\mathrm{click}}$ is given by a weighted average over this sequence with weights determined by the infinitesimal jump probability, $\delta p_{\mathrm{J}} \propto \Gamma_\text{om}(t_{\mathrm{J}})dt$ in an interval $dt$ at time $t_\text{J}$. The effects of detector dark counts as well as the optical pump filter in our experiment are incorporated via appropriate modifications to these weights.

\begin{figure}
\includegraphics[width=\columnwidth,scale=1]{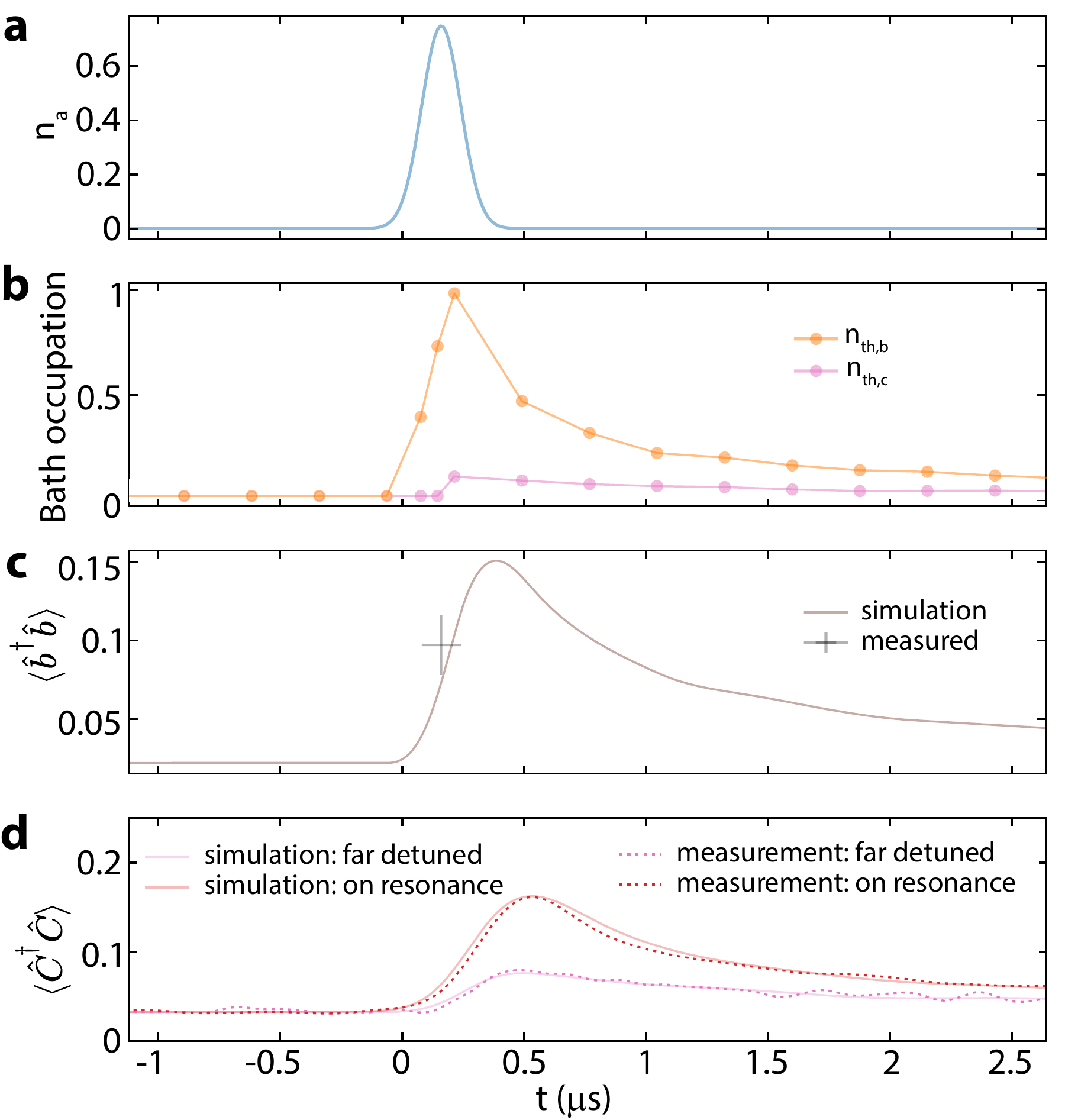}
\caption{Simulation of transducer heating dynamics \textbf{a.} Time trace of the optical pump pulse plotted in terms of the intra-cavity occupation of the optical mode, $n_a$. \textbf{b.} Model used for time dependent thermal occupation of the intrinsic baths of the acoustic and microwave modes, $n_{\mathrm{th}, b}(t)$,  $n_{\mathrm{th}, c}(t)$. \textbf{c.} Time dependent occupation of the acoustic mode, $\langle\hat{b}^\dag \hat{b} \rangle$ obtained from the master equation simulation. Gray data point corresponds to experimentally recorded thermal occupation of the acoustic mode as inferred from an optomechanical sideband asymmetry measurement. The vertical bar on the point represents a confidence interval spanning two standard deviations while the horizontal bar indicates the FWHM duration of the optical pulse. \textbf{c.} Time dependent occupation of the temporal mode, $\langle\hat{C}^\dag \hat{C} \rangle$ in the microwave output port of the transducer from the master equation simulation (solid traces) and from experimental heterodyne data (dotted traces). The condition where the acoustic mode is far detuned (on resonance) with respect to the microwave mode is shown with pink (red) traces.}
\label{si:bath_sim}
\end{figure}

\begin{figure}
\includegraphics[width=\columnwidth,scale=1]{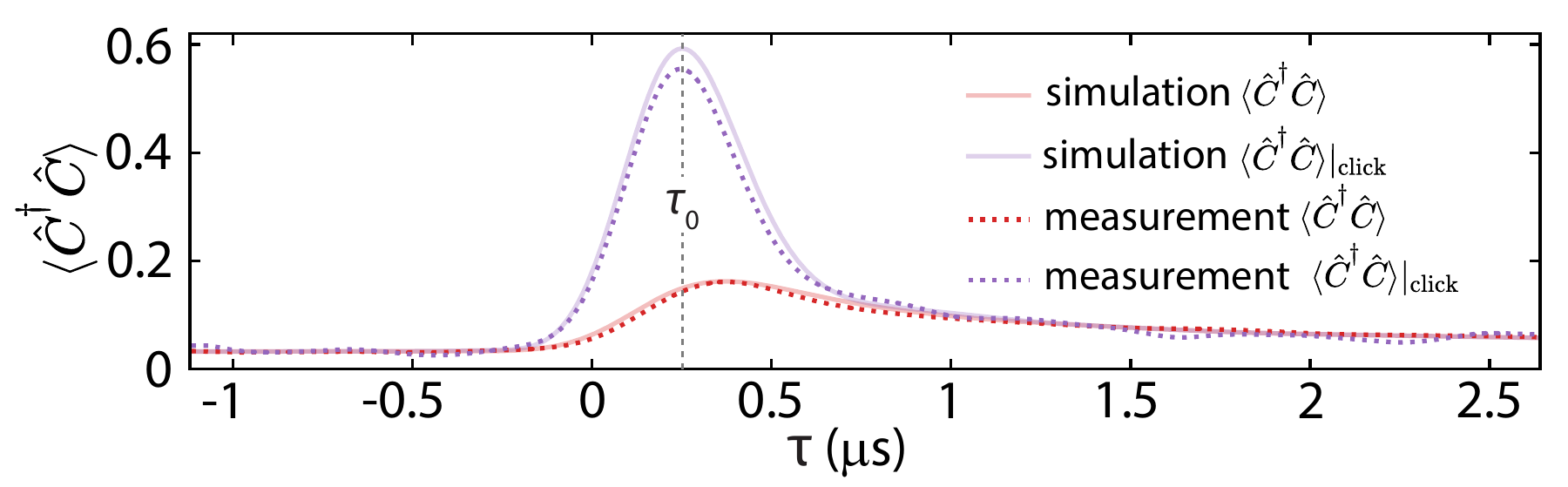}
\caption{Simulation of the conditional microwave state. Occupation of the temporal mode, $\hat{C}$ in the microwave output port of the transducer from simulation (solid traces) and from experimental heterodyne data (dotted traces). The occupation of the microwave state conditioned on an optical click, $\langle\hat{C}^\dag \hat{C} \rangle |_{\mathrm{click}}$ is shown with purple traces. The occupation of the unconditional microwave state, $\langle\hat{C}^\dag \hat{C} \rangle$ is shown with red traces for comparison.}
\label{si:g2ac_sim}
\end{figure}

We perform a numerical simulation of our system by implementing the above model using the QuTiP software package \cite{Johansson2013}. Solutions to the master equation as well as the required correlators are evaluated in a $10\times10$ Fock space of the acoustic and microwave modes. To incorporate effects of heating of the microwave and acoustic baths separately in our model, we measured the unconditional microwave emission from the transducer in response to optical pump pulses under two conditions (i) mechanics far detuned from the microwave mode, $\omega_c - \omega_b \gg 2 g_{\text{pe}}$, (ii) mechanics on resonance with the microwave mode, $\omega_b = \omega_c$. The results of the measurement are shown with the dotted traces in Fig.~\ref{si:bath_sim}d. Under condition (i), which is achieved at zero magnetic field, the microwave emission is approximately entirely due to heating of the microwave bath alone. On the other hand, condition (ii) corresponds to maximal electro-mechanical hybridization and is relevant to the SPDC experiments described in the main text. In this setting, heating of both microwave and acoustic baths is expected to contribute to the measured microwave output signal. Using measurements of the microwave emission under both conditions, we invert the master equation to determine $n_\text{th,b}(t)$ and $n_\text{th,c}(t)$ approximated as a piece-wise linear function over coarse samples as shown in Fig.~\ref{si:bath_sim}b. The inversion is performed by choosing an ansatz for $n_\text{th,b}(t)$ and $n_\text{th,c}(t)$, iteratively solving the master equation and performing least mean square optimization with respect to the experimentally measured thermal microwave emission. The resulting occupations of the acoustic mode, $\hat{b}$ and of the unconditional temporal microwave mode, $\hat{C}$ obtained by solving the master equation are shown in Fig.~\ref{si:bath_sim}c and with the solid traces in Fig.~\ref{si:bath_sim}d respectively. While the model for the hot baths is determined entirely using microwave measurements, we observe that the simulated occupation of the acoustic mode is in reasonable agreement with an experimental measurement shown by the gray data point in Fig.~\ref{si:bath_sim}c. This data point corresponds to the optomechanical sideband asymmetry measurement described in Section 6, and represents the thermal occupation of the acoustic mode averaged over the duration of the optical pump pulse. Finally, the conditional microwave emission determined using our model shown with the solid purple trace in Fig.~\ref{si:g2ac_sim} is in good agreement with the corresponding experimental result shown via the dotted purple trace. The simulated conditional and unconditional intensity traces plotted in this figure predict a value of $g^{(2)}_{AC}(\tau_o) = 3.97$, which is in good agreement with the experimentally observed value of 3.90.

Simulation of $g^{(2)}_{CC}|_{\text{click}}$ requires evaluation of $\langle\hat{C}^{\dag 2} \hat{C}^2\rangle|_{\text{click}}$. Extending the approach of Eq.~(\ref{eq:CdagC}) to the second order moment is a nontrivial computational endeavor as it requires calculation of the four-time correlator, $ \langle \hat{c}^\dag (t+t'''') \hat{c}^\dag (t+t''') \hat{c} (t+t'') \hat{c} (t+t') \rangle$. For the purpose of this work, we pursue the more tractable task of placing bounds on $g^{(2)}_{CC}|_{\text{click}}$. First, we consider the conditional state of the acoustic mode immediately after the addition of a phonon as triggered by an SPDC event. Since the extraction of this acoustic state into the microwave output port is accompanied by loss and addition of noise, we expect the function, $g^{(2)}_{bb}|_{\text{click}} = \langle \hat{b}^{\dag 2}b^2 \rangle|_{\text{click}}/(\langle \hat{b}^{\dag}b \rangle|_{\text{click}} )^2$ to provide a lower bound for $g^{(2)}_{CC}|_{\text{click}}$. Our model estimates $g^{(2)}_{bb}|_{\text{click}}(0) = 0.24$. Next, we consider a choice of the emission envelope function, $f(t) = \delta(t-T_o-\tau_o)$, where $T_o$ is the time corresponding to the peak intensity of the pump pulse. Physically, employing this function results in collecting the microwave emission at all frequencies from the transducer at a fixed time. This is strictly less optimal than the choice in our measurements, namely a function coherently matched to the theoretically expected single photon wavepacket as discussed in Sec. \ref{si:MW_EM_Section}. In the event of this sub-optimal choice,  moments of the temporal mode, $\hat{C}$ are proportional to the corresponding moments of the internal microwave mode, $\hat{c}$ and the function, $g^{(2)}_{cc}|_{\text{click} }= \langle \hat{c}^{\dag 2}c^2 \rangle|_{\text{click}}/(\langle \hat{c}^{\dag}c \rangle|_{\text{click}}) ^2$, serves as an upper bound for $g^{(2)}_{CC}|_{\text{click}}$. Our model estimates $g^{(2)}_{cc}|_{\text{click}}(\tau_o) = 0.77$. In summary, the above arguments allow us to use numerical simulations to bound $g^{(2)}_{CC}|_{\text{click}}(\tau_o)$ to the interval (0.24, 0.77), which has substantial overlap with the experimental observation of $g^{(2)}_{CC}|_{\text{click}}(\tau_o)=0.42^{+0.27}_{-0.28}$.


%
\section{Data analysis for correlation functions}
For measurements of $g^{(2)}_{AC}$ and $g^{(2)}_{CC}|_\text{click}$ presented in figure \,3 of the main text, $9.1\times10^4$ heterodyne voltage traces were acquired over a one month period. For every $\sim$16 minutes of acquisition of conditional microwave data, we interleaved the acquisition of voltage samples of amplifier noise and the unconditional microwave signal. The unconditional microwave intensity signal, $\bar{C}_{11}(\tau)$ used to evaluate the normalized cross-correlation function, $g^{(2)}_{AC}$ was obtained from $1.4\times10^7$ heterodyne voltage traces. Likewise, the measurement of the normalized second order intensity correlation function, $g^{(2)}_{CC} (\tau_\mathrm{o})$ used $3.2\times 10^7$ heterodyne voltage samples. 

\begin{figure}[!h]
\includegraphics[width=\columnwidth,scale=1]{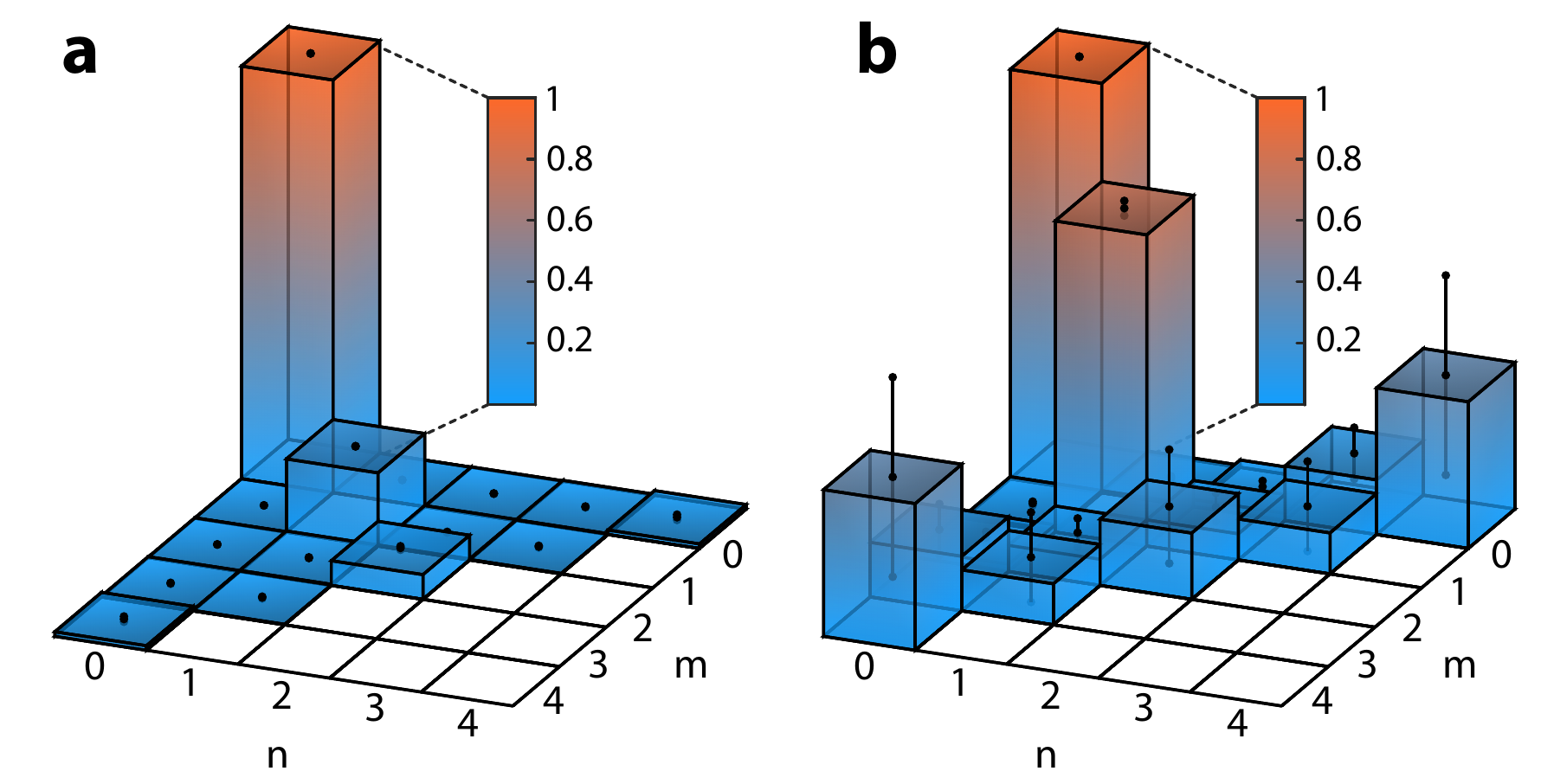}
\caption{\textbf{a.} Moments matrix for the unconditional microwave output of the transducer,  $\bar{C}_{mn}$. \textbf{b.} Moments matrix for the microwave output of the transducer conditioned on receipt of an optical click,  $\bar{C}_{mn}|_\text{click}$. Error bars calculated via bootstrapping with replacement are indicated by the black pins.}
\label{si:moments_fig}
\end{figure}

To minimize the effect of long term fluctuations in the added noise and gain of the TWPA, we divide the dataset into chunks corresponding to acquisition over one day, and invert each separately to extract the moments $\bar{C}_{mn}^{(k)}, \bar{C}_{mn}^{(k)}|_\text{click}$, where the index $k$ runs over the daily chunks. This inversion process follows the methods detailed in Sec.~\ref{sec: moment inversion} and assumes that no correlations exist between $\hat{H}$ and $\hat{A}$ as well as $\hat{H}$ and $\hat{C}$. We then take a weighted average to compute the entries of the moments matrix, $\bar{C}_{mn} = \sum_k w^{(k)} \bar{C}_{mn}^{(k)}$ and $\bar{C}_{mn}|_\text{click} = \sum_k w^{(k)}_\text{click} \bar{C}_{mn}^{(k)}|_\text{click}$, where $w^{(k)}$,  $w^{(k)}_\text{click}$ denote the fraction of records in the $k^\text{th}$ chunk of the unconditional and conditional datasets respectively. The moments matrices constructed from this process are shown in Fig.~\ref{si:moments_fig}b.

\begin{figure}[!h]
\includegraphics[width=\columnwidth,scale=1]{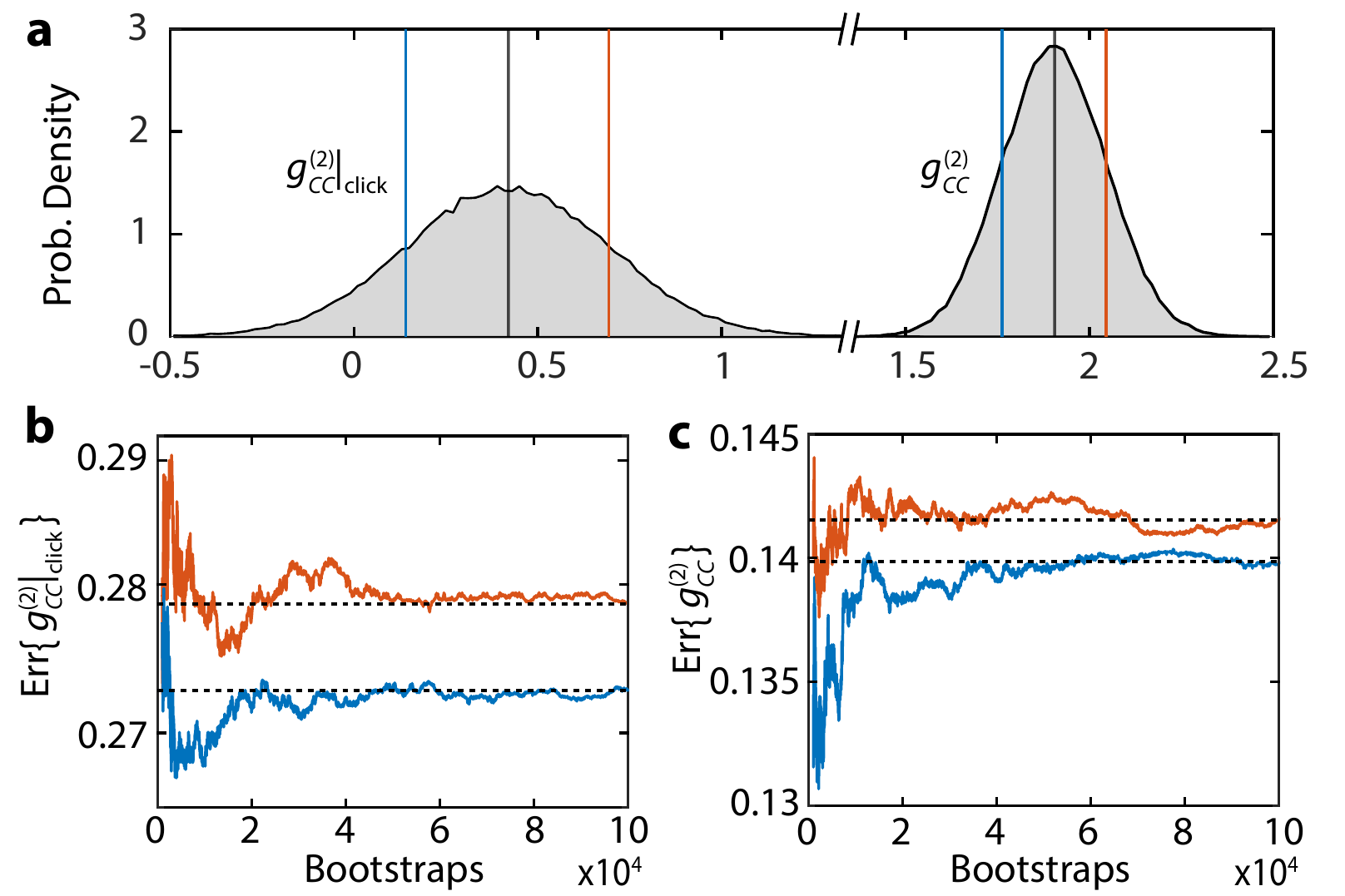}
\caption{\textbf{a.} Probability distribution for $g^{(2)}_{CC}$ and $g^{(2)}_{CC}|_\text{click}$ calculated from bootstrapping. Vertical blue and red lines correspond to one standard deviation above and below the distribution mean respectively. \textbf{b,c.} Convergence of the error estimate vs. number of bootstraps for all traces, and optically heralded traces respectively. Blue lines correspond to positive error, red lines to negative error, and black lines indicate the error estimate used in the main text.}
\label{si:convergence_fig}
\end{figure}

The acquisition rate for the moments of the conditional heterodyne output, $\bar{S}_\text{mn}|_\text{click}$ is determined by the optical heralding rate, $R_{\mathrm{click}}=0.14$ Hz. This is much slower compared to that for the amplifier noise moments, $\bar{H}_\text{mn}$, which can be acquired at the 50~kHz repetition rate of the experiment and determined with high accuracy. As a result, the error in $g^{(2)}_{CC}|_\text{click}$ is dominated by uncertainty in $\bar{S}_\text{mn}|_\text{click}$. To calculate error in $g^{(2)}_{CC}$ and $g^{(2)}_{CC}|_\text{click}$, we employ bootstrapping with replacement using $10^5$ bootstraps to construct the probability density functions shown in Fig.~\ref{si:convergence_fig}a. Finally, error estimates were calculated by numerical integration of the distributions such that they they cover a 34.1\% confidence interval above and below the mean. The error estimates calculated using this process vs. number of bootstraps is shown in Fig.~\ref{si:convergence_fig}b-c for the conditional and unconditional datasets. We find that the use of $10^5$ bootstraps yields acceptable convergence in the error estimate.

To mitigate the effects of long term drifts in frequencies of the microwave modes, we periodically measured the unconditional microwave power spectrum and monitored the resonance frequencies, $\omega_\pm$. This additionally allowed us to characterize the spectral diffusion of the microwave modes. In our data analysis, we excluded intervals of the measurement where the mode  frequencies drifted by more than twice the standard deviation associated with spectral diffusion, which represented $3\%$ of all recorded traces.

%
\section{Classical bound on the conditional second order intensity correlation}

In the main text, we claim that the conditional microwave autocorrelation function is expected to satisfy the inequality, $g^{(2)}_{CC}|_{\mathrm{click}} \geq 1$ for classical microwave-optical states. This can be proved by rewriting the conditional quantities defined in the main text in terms of correlators of the joint microwave-optical state,
\begin{align}\label{si:jointprop_eq}
    g^{(2)}_{CC}(\tau)|_\text{click} 
    &= \frac{\braket{\hat{C}^{\dagger 2}(\tau) \hat{C}^2(\tau)} |_\text{click}}
    {\left( \braket{\hat{C}^\dagger(\tau) \hat{C}(\tau)}|_\text{click}\right)^2 } \nonumber \\
    &= \frac{\braket{\hat{A}^\dagger\hat{C}^{\dagger 2}(\tau) \hat{C}^2(\tau) \hat{A}} \braket{\hat{A}^\dagger\hat{A}}}
    { \left(\braket{\hat{A}^\dagger\hat{A}} \braket{\hat{C}^\dagger(\tau) \hat{C}(\tau)} \right)^2},
\end{align}
where the above correlators are explicitly written in normal order. Using the the optical equivalence theorem, we can recast the expectation values of the above correlators in phase space using the Sudarshan-Glauber P representation \cite{Mandel1995,Kuzmich2003} as
\begin{align}
    \braket{\hat{A}^{\dagger m} \hat{C}^{\dagger n}(\tau) \hat{A}^m \hat{C}^n(\tau) } 
    & = 
    \int|\alpha|^{2m} |\gamma|^{2n} P_\tau(\alpha,\gamma)  d^2 \alpha d^2\gamma \nonumber \\
    & :=  \braket{|\alpha|^{2m} |\gamma|^{2n}}_{P_\tau},
\end{align}
where $P_\tau(\alpha,\gamma)$ is the joint phase space density cprresponding to modes $\hat{A}$ and $\hat{C}$ with relative delay $\tau$. Applying this directly to Eq.~(\ref{si:jointprop_eq}) gives,
\begin{equation}
    g^{(2)}_{CC}(\tau)|_\text{click}  = \frac{\braket{|\alpha|^2_P |\gamma|^4}_{P_\tau} \braket{|\alpha|^2}_{P_\tau}}{\braket{|\alpha|^2|\gamma|^2}^2_{P_\tau}} \geq 1,
\end{equation}
which follows directly from the Cauchy-Schwarz inequality.


%
\section{Convergence of $g^{(2)}_{AC}$ to the classical bound}
\begin{figure}[!h]
\includegraphics[width=\columnwidth,scale=1]{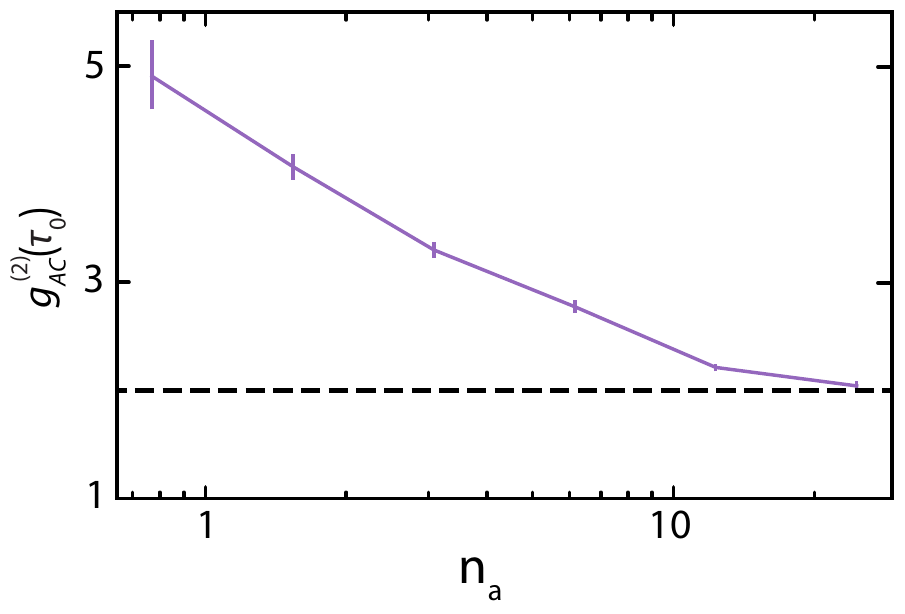}
\caption{Maximum value of $g^{(2)}_{AC}$ versus increasing pump power plotted on the horizontal axis as the peak intra-cavity occupation of the optical mode, $n_a$. Dashed line denotes the classical upper bound of 2.}
\label{si:cc_powerdep}
\end{figure}

The results of pump power dependent measurements of the normalized intensity cross-correlation function, $g^{(2)}_{AC}$ are shown in Fig.~\ref{si:cc_powerdep}. 

All measurements were performed with the same pump pulse duration and repetition rate used to acquire data in the main text. We observe that at higher pump powers, the value of $g^{(2)}_{AC}$ approaches the classical upper bound of 2 as expected from the increased thermal noise added to the photon pair generated in SPDC. The data in this figure was collected from the same device used for the experiments in the main text albeit after a partial warm up and cooldown to base temperature. This thermal cycle led to modified device parameters, which produced an increase in $g^{(2)}_{AC}$ at the lowest pump power of $n_a=0.8$ compared to the result presented in the main text.

\bibliography{refs}